\documentclass[instruments,article,accept,pdftex,moreauthors]{Definitions/mdpi}

\usepackage{amsmath}
\usepackage{amssymb}
\usepackage{graphicx}
\usepackage{subfig}
\usepackage{multirow,booktabs}
\usepackage{siunitx}
\usepackage{geometry}

\newcommand{\apj}{ApJ}
\newcommand{\apjl}{ApJ Lett.}

\newcommand{\solphys}{Sol. Phys.}
\newcommand{\mnras}{MNRAS}

\newcommand{\pasj}{PASJ}

\firstpage{1}

\pubvolume{10}
\issuenum{1}
\articlenumber{21}
\pubyear{2026}
\copyrightyear{2026}
\externaleditor{Martin Pohl}
\datereceived{13 February 2026}
\daterevised{22 March 2026}
\dateaccepted{30 March 2026}
\datepublished{3 April 2026}

\Title{Observational Technological Innovations and Future Development of the Lijiang Coronagraph}

\Author{Xuefei Zhang
 $^{1,2}$\orcidA{}, Yu Liu $^{3,}$*\orcidB{}, Tengfei Song $^{1,2}$\orcidC{}, Mingyu Zhao $^{1,2}$, Xiaobo Li $^{1}$, Mingzhe Sun $^{4}$, Feiyang Sha $^{3}$ and Xiande Liu $^{1,5}$}

\AuthorNames{Xuefei Zhang, Yu Liu, Tengfei Song, Mingyu Zhao, Xiaobo Li, Mingzhe Sun, Feiyang Sha and Xiande Liu}

\address{
$^{1}$ \quad Yunnan Observatories, Chinese Academy of Sciences, Kunming 650216, China;\linebreak {zhangxuefei@ynao.ac.cn} 
 (X.Z.); stf@ynao.ac.cn (T.S.); myzhao@ynao.ac.cn (M.Z.); lixiaobo@ynao.ac.cn X.L.)
\\
$^{2}$ \quad Yunnan Key Laboratory of Solar Physics and Space Science, Kunming 650216, China\\
$^{3}$ \quad College of Physical Science and Technology, Southwest Jiaotong University, Chengdu 611756, China; shafy@my.swjtu.edu.cn\\
$^{4}$ \quad Shandong Provincial Key Laboratory of Optical Astronomy and Solar-Terrestrial Environment, Institute of Space Sciences, Shandong University, Weihai 264209, China; sunmingzhe2003@126.com\\
$^{5}$ \quad School of Physics and Astronomy, Yunnan University, Kunming 650504, China; liuxiande@ynao.ac.cn\\
}

\corres{Correspondence: lyu@swjtu.edu.cn }

\abstract{As a core ground-based coronal observation facility in the low-latitude and high-altitude regions of China, the Lijiang Coronagraph takes advantage of the natural endowments of the Lijiang Astronomical Observation Station, such as an altitude of 3200 m and low atmospheric turbulence. It has gone through a complete development process from introduction through Chinese--Japanese cooperation to independent innovation and iteration. This paper systematically summarizes the core technological innovation achievements of this facility, including the upgrade of the automatic operating system, the integration of the dual-band observation system, the stray light suppression technology based on the image difference method before and after cleaning, and the high-precision image calibration and registration technology. These innovations have significantly improved observation efficiency and data quality, laying a solid foundation for high-quality observations. At the scientific research level, the observation data reveal that 1.1~$R_\odot$ (solar radius) is a highly correlated region between coronal green line brightness and magnetic field intensity. This study also confirms a strong correlation between the coronal green line and the SDO/AIA 211~\AA\ extreme ultraviolet band (correlation coefficient: 0.89--0.99), which can support the research on early warning of Coronal Mass Ejections (CMEs). These achievements provide key data support for the verification of coronal heating mechanisms and the exploration of the origin of the slow solar wind. The technical experience accumulated from the Lijiang Coronagraph has not only laid a solid foundation for the research and development of China's  next-generation large-aperture coronagraphs, but also facilitated and accelerated substantial progress in China’s technical capabilities for low coronal observation, enabling the country to establish internationally parallel competitive capabilities in this field. This system has also become an important part of the global coronal observation network. }


\begin{document}

\section{Introduction}
\label{sec:introduction}

The plasma dynamic evolution process of the corona, as~the outermost layer of the solar atmosphere, is closely related to major frontier issues in solar physics, such as coronal heating mechanisms, solar wind origin, and CMEs. It is a core research object for revealing the laws of solar activities and space weather effects \citep{wang1997ApJ, yang2024Sci, Chen2023MNRAS, song2025ApJ}. Since the brightness of the corona is only one millionth of that of the solar photosphere, conventional observations are severely interfered with by strong light. Ground-based coronagraphs, through artificial solar eclipse technology, have broken through the natural limitation of short-term observations during total solar eclipses and become the core supporting equipment for regular coronal observations~\citep{Lyot1939MNRAS, Tian2013SoPh, Wijn2012SPIE, Liang2021MNRAS}. High-altitude regions have unique advantages such as weak atmospheric turbulence, low background brightness, and~a large number of sunny days per year, making them the preferred areas for the construction of ground-based coronagraphs. They provide natural observation conditions for capturing fine coronal structures and weak radiation signals~\citep{han2021MNRAS, rao2016ApJ}.

The Lijiang Astronomical Observation Station in Yunnan (altitude: 3200~m, longitude: 100$^\circ$2$^\prime$ E, latitude: 26$^\circ$42$^\prime$ N) has become the primary site for low-latitude and high-altitude coronal observation in China due to its excellent observation environment~\citep{Wu2014AcASn}. {The Yunnan Observatories Coronagraph Green-line Imaging System (YOGIS) is a 100 mm aperture internal-occulting coronagraph deployed at Lijiang Observatory. Developed through a collaborative effort between the Yunnan Observatories and the National Astronomical Observatory of Japan (NAOJ), this instrument is specifically engineered to observe the coronal green line within the radial range from 1.03 to 2.5 \mbox{$R_\odot$}. YOGIS represents an enhanced iteration of the NOrikura Green-line Imaging System (NOGIS), which was initially constructed by NAOJ}~\citep{ichinoto1999PASJ, sakurai2012ASPC}. As~China's first ground-based internally occulted coronagraph put into regular operation, its completion has filled the gap in the field of continuous low-latitude coronal observation in China. However, the~design parameters of the NOGIS were optimized for its original observational environment and scientific objectives. After~its relocation to Lijiang, the~instrument is confronted not only with challenges posed by the high-altitude, low-pressure conditions, the~updated scientific objectives, and~the upgrade of the integrated system, but~also with the demand for high-time-resolution observations of solar eruptive events during the solar activity maximum. Its original observational environment, detector performance, and~operating system are no longer sufficient to meet the coronagraph’s requirements for high signal-to-noise ratio, high resolution, and~integrated multi-device observations \citep{song2025Univ}.

To this end, we have carried out systematic technical upgrades and multi-channel observation system integration for the YOGIS, with~the central objectives of adapting it to the extreme high-altitude observational environment in Lijiang, improving its temporal resolution, and~enhancing the instrument's autonomous observation capability. This paper focuses on elaborating the key technological breakthroughs of YOGIS in the optimization of the core optical system, the~iteration of detection devices, the~innovation of observation modes, and~the construction of dual-band observation channels. It also systematically analyzes the performance improvement of the upgraded instrument in detecting fine coronal structures and capturing dynamic characteristics of plasma. These research achievements not only provide high-quality observation data support for the study of basic scientific issues such as coronal heating mechanisms and solar wind origin, but~also offer reusable technical paradigms and engineering practice bases for the independent research and development as well as multi-band upgrade and transformation of subsequent ground-based coronagraphs in China. They are of great strategic significance for promoting China's low coronal observation to leap from following the international pace to keeping pace with it.

\section{Development History and Technical Orientation of Lijiang Coronal~Observation}
\label{sec:lijiang_coronal_obs_history_orient}

\begin{figure}[H]

\subfloat{\includegraphics[width=6.5cm,height=6.0cm]{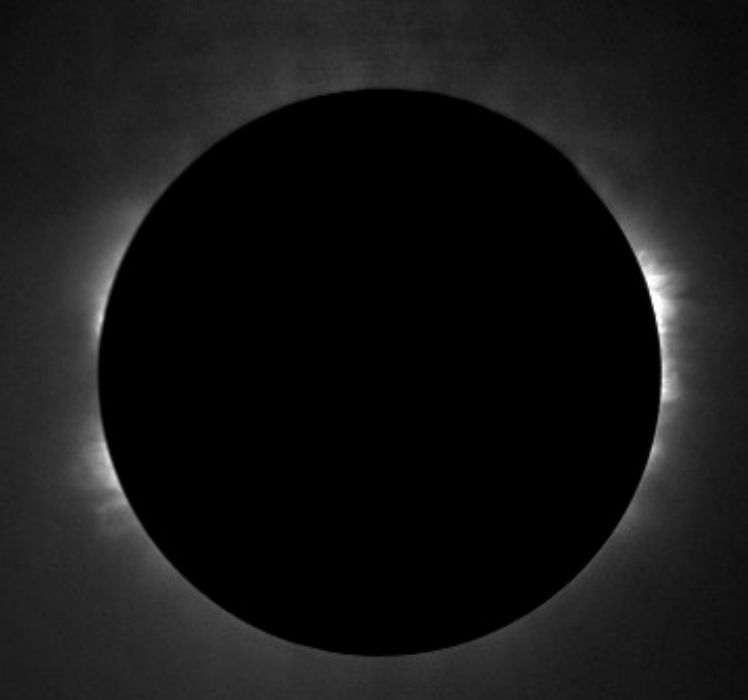}}  
\subfloat{\includegraphics[width=6.5cm,height=6.0cm]{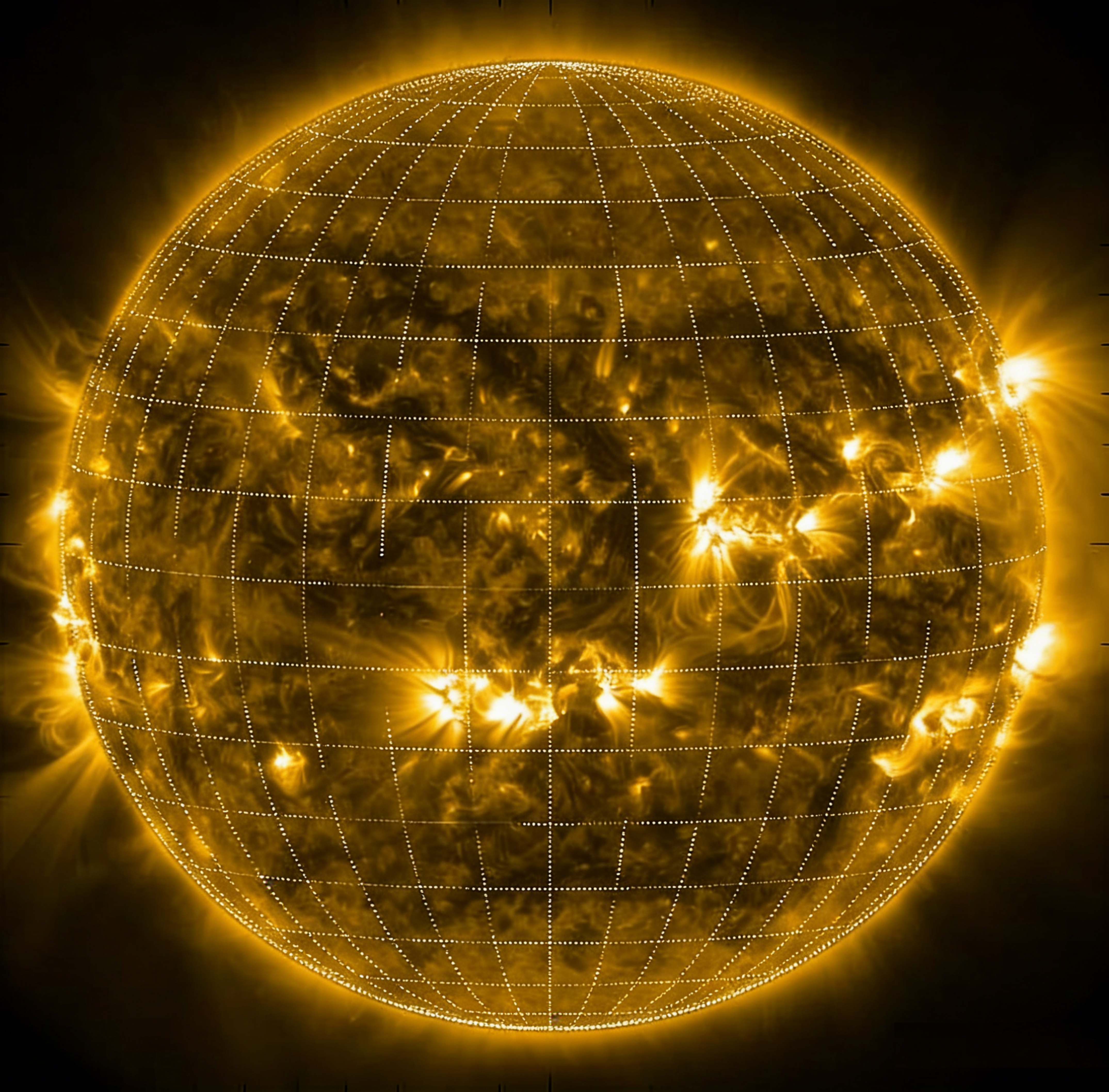}}

\caption{On 25 October {2013}, the~YOGIS successfully acquired its first coronal images. (\textbf{Left}):~visible-band data obtained by the Lijiang coronagraph; (\textbf{right}): extreme ultraviolet (EUV) band data from the Solar Dynamics Observatory (SDO) \citep{Lemen2012SoPh}.}
\label{fig1}
\end{figure}

Since 2010, the~coronagraph team of Yunnan Observatories has launched a large-scale site selection project. It has conducted systematic monitoring for more than two years in over 60 candidate regions including Xinjiang, Tibet, and~Yunnan, and~finally selected two high-quality observation sites, namely Lijiang in Yunnan and Daocheng in Sichuan~\citep{Xu2023RAA}. In~2013, we successfully built a green line coronagraph at the Lijiang Astronomical Observation Station and obtained clear coronal green line images for the first time. This has strongly confirmed the great potential of the high-altitude regions in western China in the field of coronal observation and laid a solid foundation for subsequent technological research and development (Figure~\ref{fig1}).

\vspace{-6pt}

\begin{figure}[H]
\subfloat{\includegraphics[width=4.9cm,height=6.0cm]{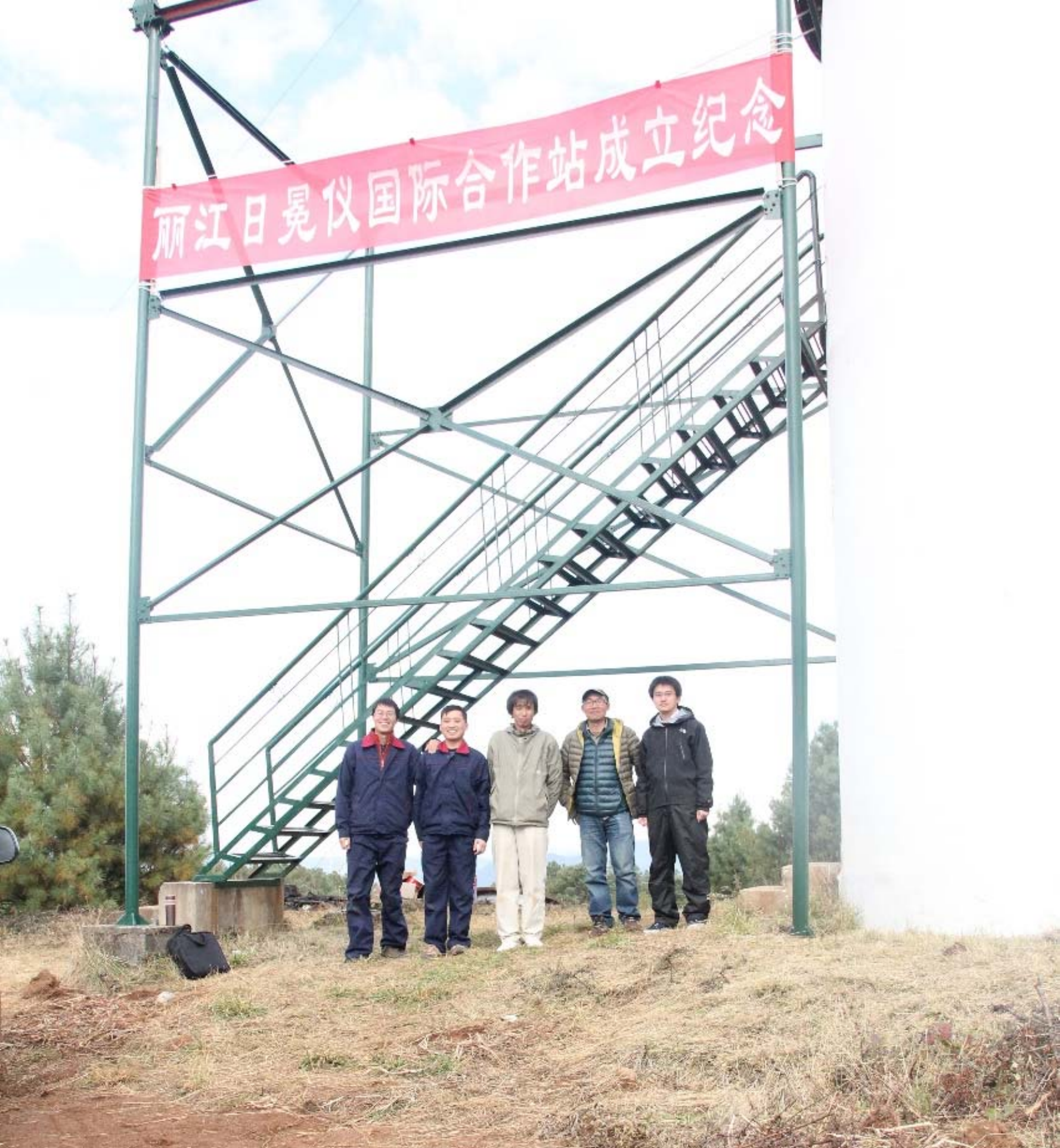}}  
\subfloat{\includegraphics[width=8.0cm,height=6.0cm]{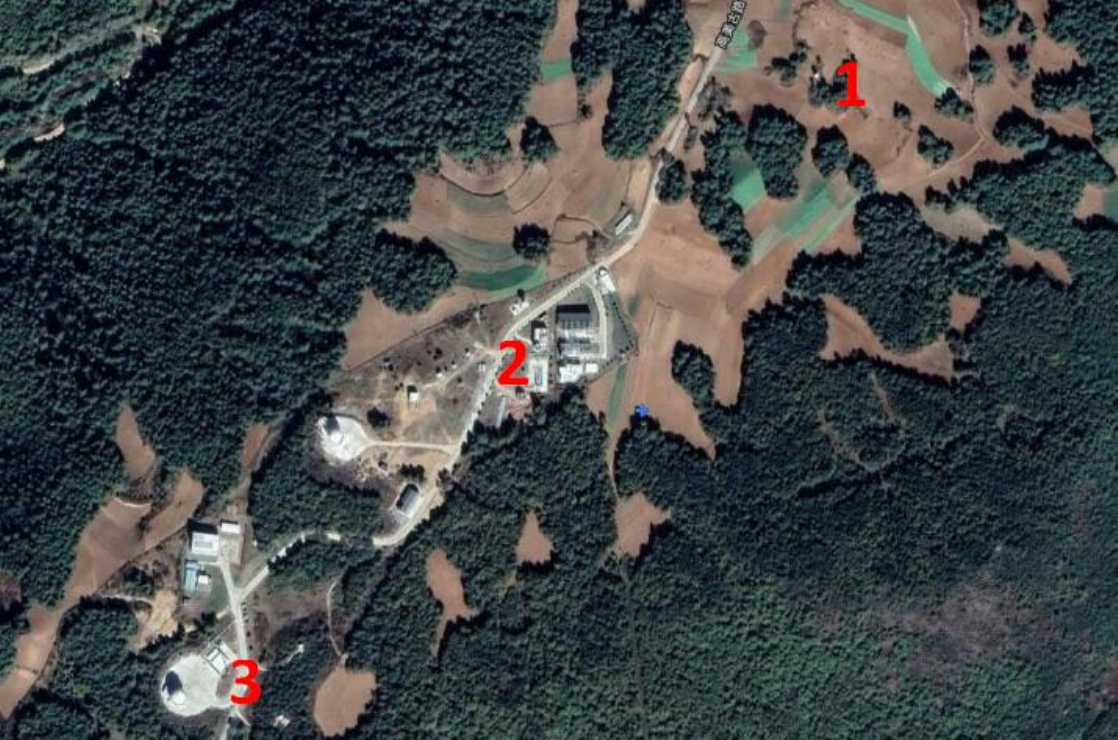}}

\caption{(\textbf{Left}): Commemorative photo of the completion of the YOGIS international cooperation station. (\textbf{Right}): location 1 indicates the original construction site of YOGIS; location 2 denotes the high-altitude test base for the coronagraph; location 3 marks the site of the Lijiang 2.4-m optical telescope . The~image is sourced from Google~Earth.}
\label{fig2}
\end{figure}

The development history of Lijiang coronal observation is a history of continuous exploration and constant breakthroughs in technological iteration.  {It is inseparable from the pioneering research and accumulated technical expertise in coronagraph observation techniques and coronal magnetic field measurement methods, conducted through our collaboration with institutions such as the Mauna Loa Solar Observatory (MLSO) and the Institute for Astronomy, University of Hawaii} \citep{Zhang2023soma, gong2004SPIE}. Since the establishment of the Chinese--Japanese cooperative coronal observation station in 2013, the~equipment has undergone three generations of technological upgrades, and~its observation capability has achieved leapfrog improvement. {The key parameters of YOGIS are summarized in Table} \ref{tab:yogis_params}. In~2017, the~team, in~cooperation with the University of Science and Technology of China, built a high-altitude experimental base for coronagraphs in Lijiang (Figure~\ref{fig2}). In~2018, a~prototype coronagraph developed under the leadership of Shandong University with the in-depth participation of this team successfully obtained coronal green line data at this base, achieving a series of key technological breakthroughs in coronagraph development. The~site construction and ground experiments during this period have accumulated a large amount of valuable measured data for core technologies such as environmental adaptability optimization and stray light suppression of balloon-borne~coronagraphs.

\begin{table}[H]

    \caption{Key characteristics of the~YOGIS.}
    \label{tab:yogis_params}
    
\begin{adjustwidth}{-\extralength}{0cm}
\begin{tabularx}{\fulllength}{m{3.5cm}<{\raggedright}m{4.3cm}<{\raggedright}m{5.2cm}<{\raggedright}m{4cm}<{\raggedright}}
        \toprule
        \textbf{Category} & \textbf{Parameter} & \textbf{Specification} & \textbf{Notes} \\
        \midrule
        \multirow{4}{*}{Optical System} & Instrument Type & Internally occulted Lyot coronagraph & NOGIS-based \\
        & Objective Lens Aperture & 100 mm & Original NOGIS spec. \\
        & Focal Length & 1490 mm (at 5303~\AA) & Green line optimized \\
        & Field of View & $1.03$--$2.5~R_{\odot}$ & Inner corona \\
        \midrule
        \multirow{6}{*}{\vspace{-18pt}Filter \& Detection} & Primary Observing Wavelengths & Fe XIV 5303~\AA~(green) \& Fe X 6374~\AA~(red) & Red line added \\
        & Spectral Bandwidth & 1~\AA\ FWHM & Lyot filter \\
        & Filter System & Four-stage birefringent Lyot filter with LCVR tuning & Fast switching \\
        & Wavelength Tuning Range & $\pm2$~\AA\ around each line & Doppler capable \\
        & Tuning Speed & <60 ms & LCVR tuning \\
        & Detector & Cooled CMOS, $2048\times2048$ pixels & CCD upgrade \\
        \midrule
       \multirow{2}{*}{\vspace{-15pt}Observation \& Control} & Observation Modes & Standard, fast (15 s/frame), engineering; dual-band alternating & Event-driven \\
        & Autonomous Operation & ASCOM-based integrated control system & Multi-device sync \\
        \midrule
        Stray Light Suppression & Methods & Internal black coating; dust imaging path; image differencing & Dust scattering removal \\
        \midrule
        Site & Location & Lijiang Astronomical Observation Station & Altitude: 3200 m \\
        \bottomrule
    \end{tabularx}
\end{adjustwidth}

\end{table}
\unskip
\vspace{-6pt}

Relying on a sub-project of the Strategic Priority Research Program (Class A) ``Honghu Project'' of the Chinese Academy of Sciences, Yunnan Observatories, together with multiple scientific research institutions, launched the research and development of a balloon-borne white-light coronagraph. It successfully overcame key technical bottlenecks such as the suppression of stray light at the million-level and developed a 50 mm aperture white-light coronagraph~\citep{Liu2021SPIE}. On~{27 February 2021}, this instrument completed ground observations at Wuming Mountain in Daocheng, Sichuan, with~an altitude of nearly 4800 m. It obtained the first white-light coronal images through China's independently developed white-light coronagraph, fully verifying the ground observation performance of the equipment. On~{4 October 2022}, with~the help of the high-altitude balloon platform of the Aerospace Information Research Institute of the Chinese Academy of Sciences, the~coronagraph was sent to the stratosphere at an altitude of 30 km in Da Qaidam, Haixi Prefecture, Qinghai~\citep{Li2024ChJSS}. It successfully carried out continuous observations for 5 h and obtained nearly 20,000 white-light coronal photos. This is the first time in the international solar physics community that white-light observations of the inner corona have been carried out at this altitude, {demonstrating that our team has the capacity to develop and conduct similar experiments}~\citep{Gopalswamy2021SoPh, Songa2023ApJ}.

\begin{figure}[H]

\begin{adjustwidth}{-\extralength}{0cm}
\centering 
\includegraphics[width=18.0cm, height=6.0cm]{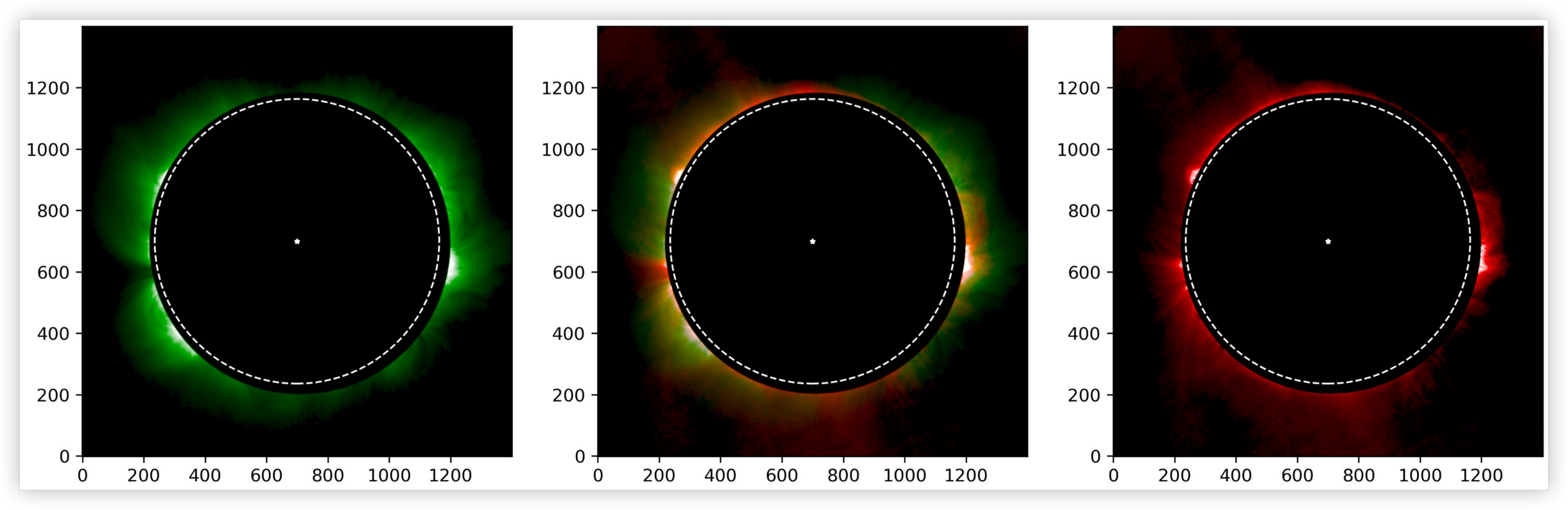}
\end{adjustwidth}

\caption{{Equipped with a multi-band filter system, the~SICG coronagraph supports dual-band coronal observations: green-line (5303~\AA, (\textbf{left})), composite green and red overlay (\textbf{middle}), and~red-line (6374~\AA, (\textbf{right})).}}
\label{fig2-1}
\end{figure}

With the continuous progress of technology, Lijiang coronal observation has ushered in a landmark upgrade. The~Spectroscopic Imaging Coronagraph for Meridian Project Phase II (SICG), as~China's first independently developed ground-based coronagraph put into regular operation, passed the process test and obtained the first batch of coronal observation images in October 2023, filling the relevant gaps in the field of coronagraph development in China \citep{Liu2025RAA}. {SICG is a key instrument constructed as part of the second phase of CMP-II, specifically designed for inner corona observations over a field of view of 1.05 to 2.0 \mbox{$R_\odot$}. It operates at two primary wavelengths---6374~\AA~and 5303~\AA---corresponding to the red and green emission lines of the corona. The~coronagraph is equipped with a 200~mm objective lens, with~focal lengths of 2000.2 mm at 6374~\AA~and 1983.3 mm at 5303~\AA. A~distinctive feature of its optical design is the use of a birefringent prism to split the light path into two branches: the reflected beam is directed toward stray light monitoring of the objective lens, while the transmitted beam passes through a tunable birefringent filter for selecting the targeted E-corona emission lines. Wavelength selection and switching are achieved through a combination of pre-filters and a four-stage Lyot filter} \citep{Tomczyk2008SoPh, Morton2016ApJ}.  {This image displays the quasi-simultaneous observation results of the coronal green line and red line} in Figure~\ref{fig2-1}.The Fe X 6374 \AA\ spectral line can be used to study the magnetohydrodynamic coupling process and material transport phenomena in the low-temperature corona, while the Fe XIV 5303\AA\ spectral line can help explore the heating mechanism and energy release process in the high-temperature corona~\citep{Yang2026ApJS, Li2017ApJ, Priyal2025ApJ}. The~collaborative observation of these two spectral lines provides powerful technical support for a comprehensive analysis of coronal physical processes, and enables Lijiang coronal observation to occupy an important position in the international coronal observation network, becoming a key node for studying the magnetohydrodynamic coupling process of the low~corona.

\unskip

\section{Observational Technological Innovations: From Noise Suppression to Precision~Calibration}
\label{sec:obs_tech_innov_noise2calib}
\unskip

\subsection{Reconstruction of the Observation~Platform}
\label{sec:obs_platform_reconstruction}

\begin{figure}[H]

\begin{adjustwidth}{-\extralength}{0cm}
\centering 
\subfloat{\includegraphics[width=7.0cm, height=6.0cm]{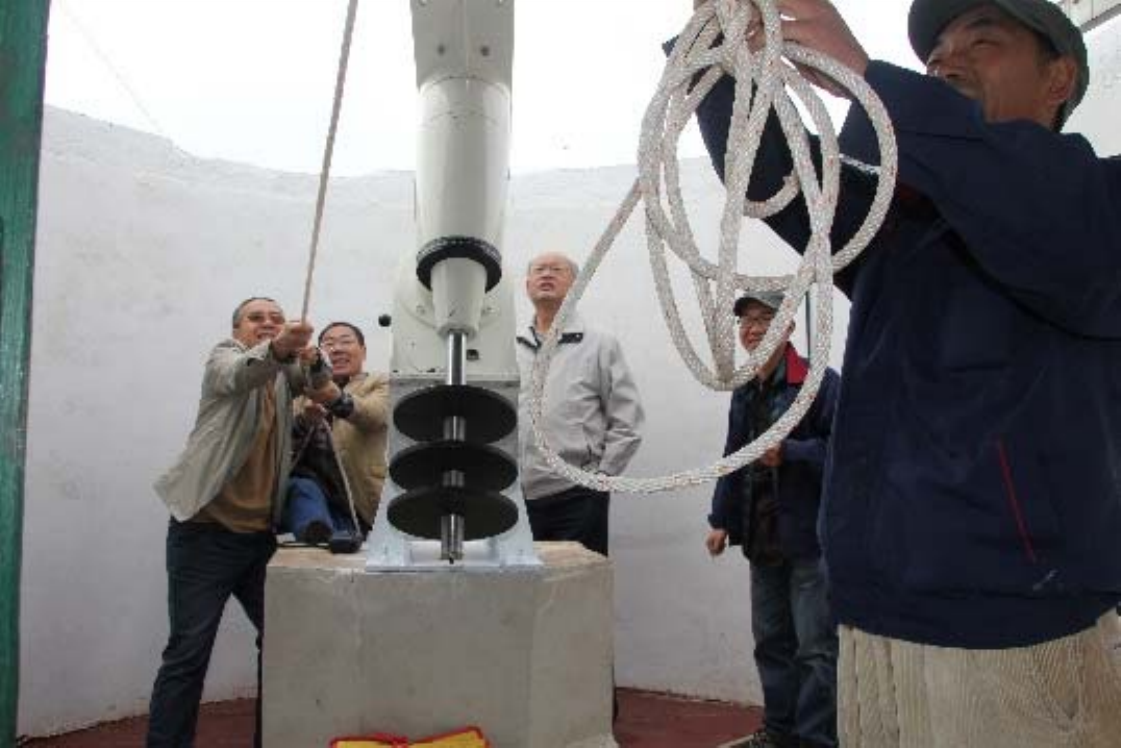}}  
\subfloat{\includegraphics[width=7.0cm, height=6.0cm]{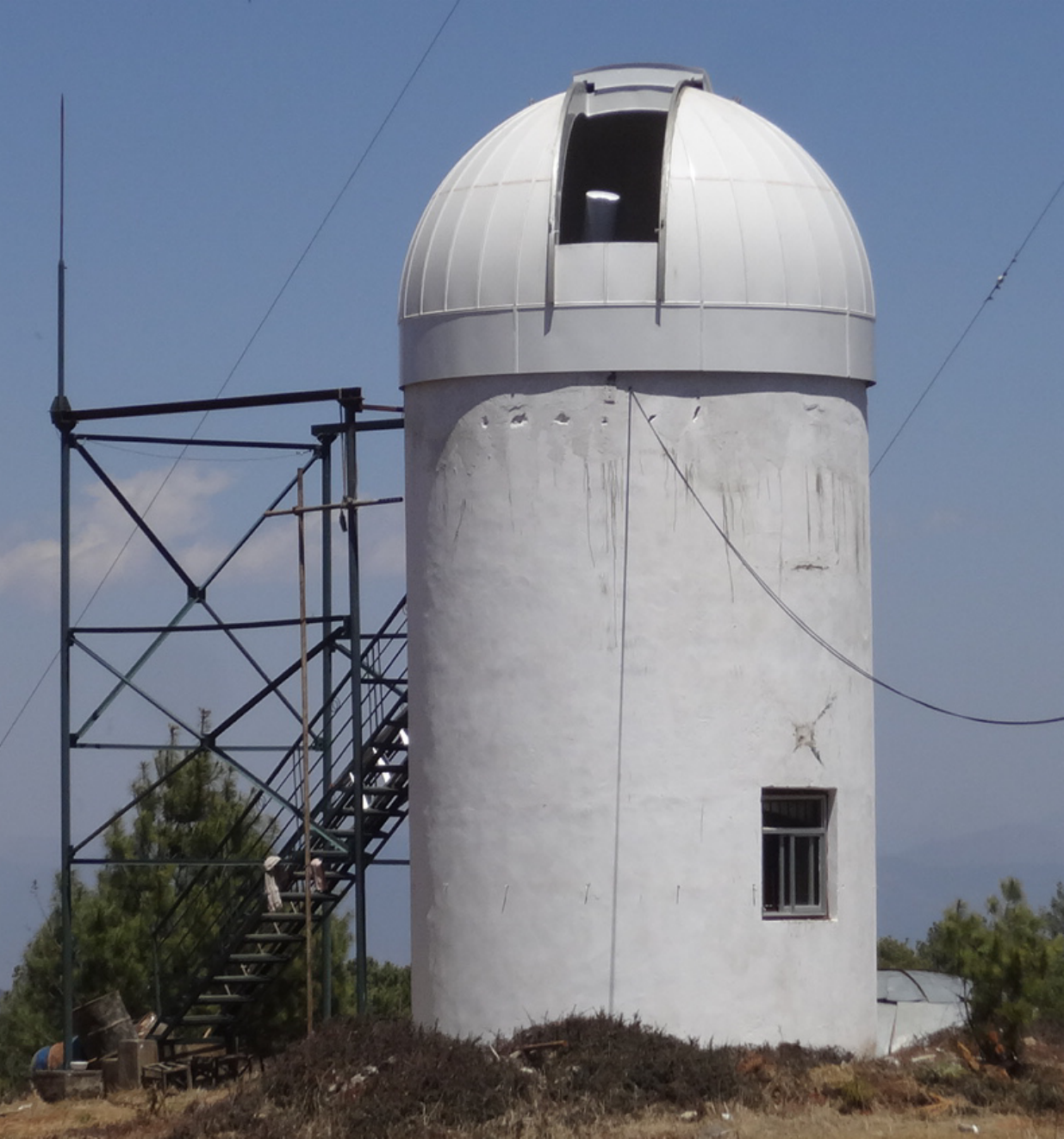}}
\end{adjustwidth}

\caption{(\textbf{Left}): In the early stages of YOGIS construction in 2013, the~coronagraph lacked rainproof facilities. A~movable rain shelter was installed on the roof of the circular building. The~image shows staff members opening the roof shelter via pulleys to conduct observations. (\textbf{Right}): In 2014, a~dome was designed and installed according to the dimensions of the circular building, upgrading the equipment for safeguarding the observation~environment.}
\label{fig3}
\end{figure}

As China's first ground-based coronagraph put into regular observation, YOGIS officially began its construction in 2013. The~initial observation site was selected with significant scientific consideration, located 1 km outside the Lijiang Astronomical Observation Station compound—a location that was once a core candidate site for monitoring the Lijiang Astronomical Observatory's foundation \citep{Tan2002BASI}. During~the initial observation phase, a~dedicated observation platform was already built at this site. The~platform featured a circular structure with a diameter of 4.5 m, erected approximately 12 m above the ground, effectively mitigating the interference of near-surface atmospheric turbulence on observations. Notably, this initial platform was not equipped with a dome facility commonly used in astronomical observations; however, the optical components of the coronagraph are extremely sensitive to environmental factors such as humidity and rain. Therefore, in~the early stages of YOGIS's operation, customized rainproof equipment was specifically installed. This setup not only protected the instrument from rain and snow erosion but also minimized the obstruction of the observation field of view caused by the equipment itself (Figure~\ref{fig3}).

\vspace{-6pt}

\begin{figure}[H]

\subfloat{\includegraphics[width=6.0cm, height=6.0cm]{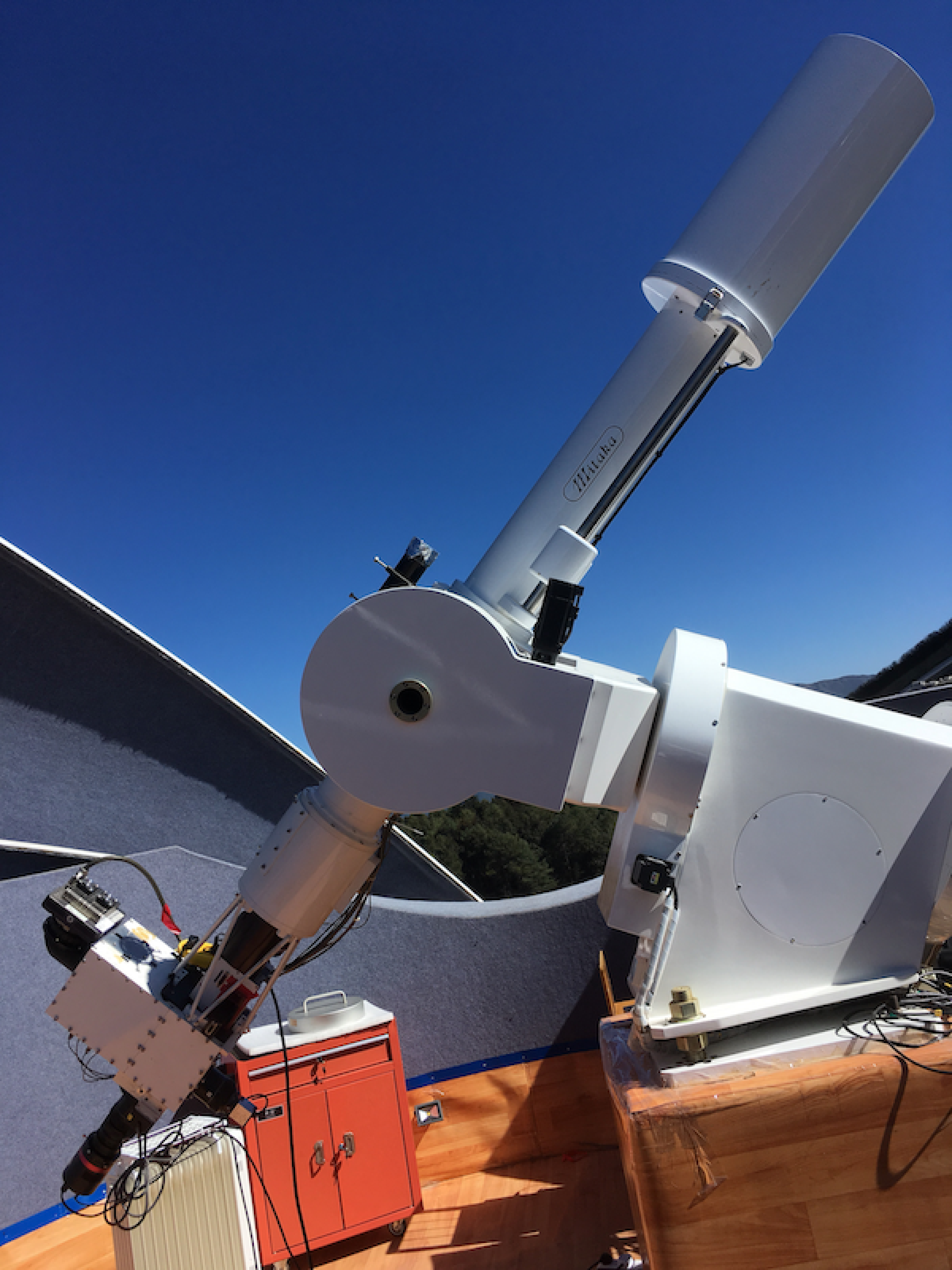}}  
\subfloat{\includegraphics[width=6.0cm, height=6.0cm]{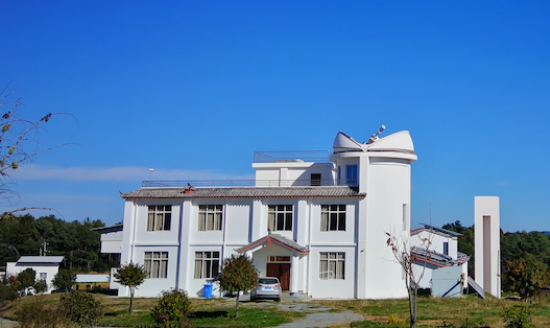}}

\caption{(\textbf{Left}): YOGIS operates on the newly constructed platform, ensuring the regular coronal monitoring capability of the coronagraph. (\textbf{Right}): This is the high-altitude coronagraph test base, dedicated to conducting ground-based coronagraph testing and experimental research; China's  independently developed regular-operation coronagraph underwent its trials right~here.}
\label{fig4}
\end{figure}

To further enhance observation stability and data quality, the~new YOGIS observation platform was officially completed and put into operation in 2017, relying on multi-party technical collaboration and joint research efforts. The~core breakthrough of the new platform lies in its pier design, with~its natural frequency successfully increased to over 12~Hz. This key indicator far exceeds the requirements for regular observations, effectively resisting external disturbances such as wind forces and ground vibrations, and~thus providing a solid structural guarantee for the high-precision tracking observation of coronal activities. Meanwhile, the~supporting high-altitude test site for the coronagraph was completed simultaneously. This test site not only provides an exclusive experimental space for YOGIS equipment upgrades and parameter optimization, but~also establishes a domestic R\&D and verification platform for key coronagraph technologies, laying an important practical foundation for the development of subsequent independently developed coronal observation instruments in China (Figure~\ref{fig4}).
\unskip

\subsection{Intelligent Upgrading of the Observation Control~System}
\label{sec:obs_control_sys_intel_upgrade}

During the initial construction phase of YOGIS, constrained by technical conditions, the~supporting graphical user interface (GUI) had relatively simple functions, only capable of displaying basic observation data, with~core operational procedures highly dependent on manual intervention. Specifically, equipment controls such as dome rotation and the start/stop of observation instruments had to be manually performed on site. Monitoring of critical environmental parameters, including atmospheric humidity, temperature, and~wind speed, also relied on manual recording and~judgment.

To address this issue, we introduced the Astronomy Common Object Model (ASCOM{, an~open-source astronomical control platform originally developed by the astronomical community in the United States, as~requested})---Alpaca open-source standardized astronomical control platform to reconstruct and upgrade the observation control system \citep{song2025Univ, ASCOM}. Through systematic integration, the~three core systems of YOGIS---the telescope system responsible for target acquisition, the~pointing and tracking system ensuring observation precision, and~the auxiliary system covering environmental monitoring and equipment power supply—were organically integrated, enabling collaborative linkage and centralized management of each subsystem. Meanwhile, based on the actual needs of scientific research and operation and maintenance (O\&M), three differentiated observation modes were innovatively designed: the Standard Observation Mode, suitable for regular scientific observation tasks, which can automatically complete the entire observation process according to preset parameters; the Rapid Observation Mode, targeting sudden solar events such as CMEs, which can quickly activate the equipment and focus on the observation target; and the Engineering Debugging Mode, which provides a dedicated interface for technicians to calibrate equipment parameters and troubleshoot faults, balancing research efficiency with O\&M~convenience.

\begin{figure}[H]

\subfloat{\includegraphics[width=5.0cm, height=6.0cm]{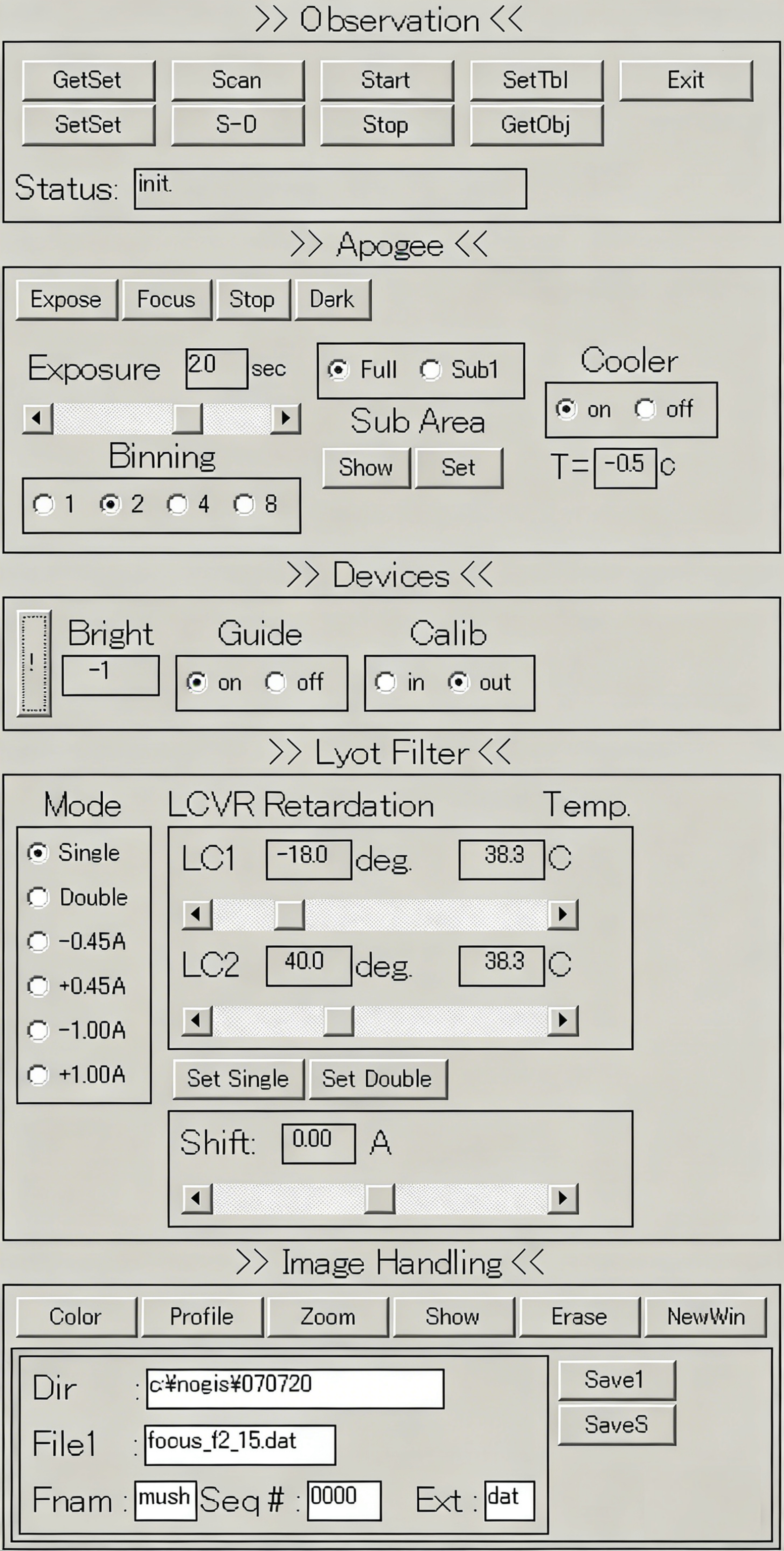}}  
\subfloat{\includegraphics[width=9.0cm, height=6.0cm]{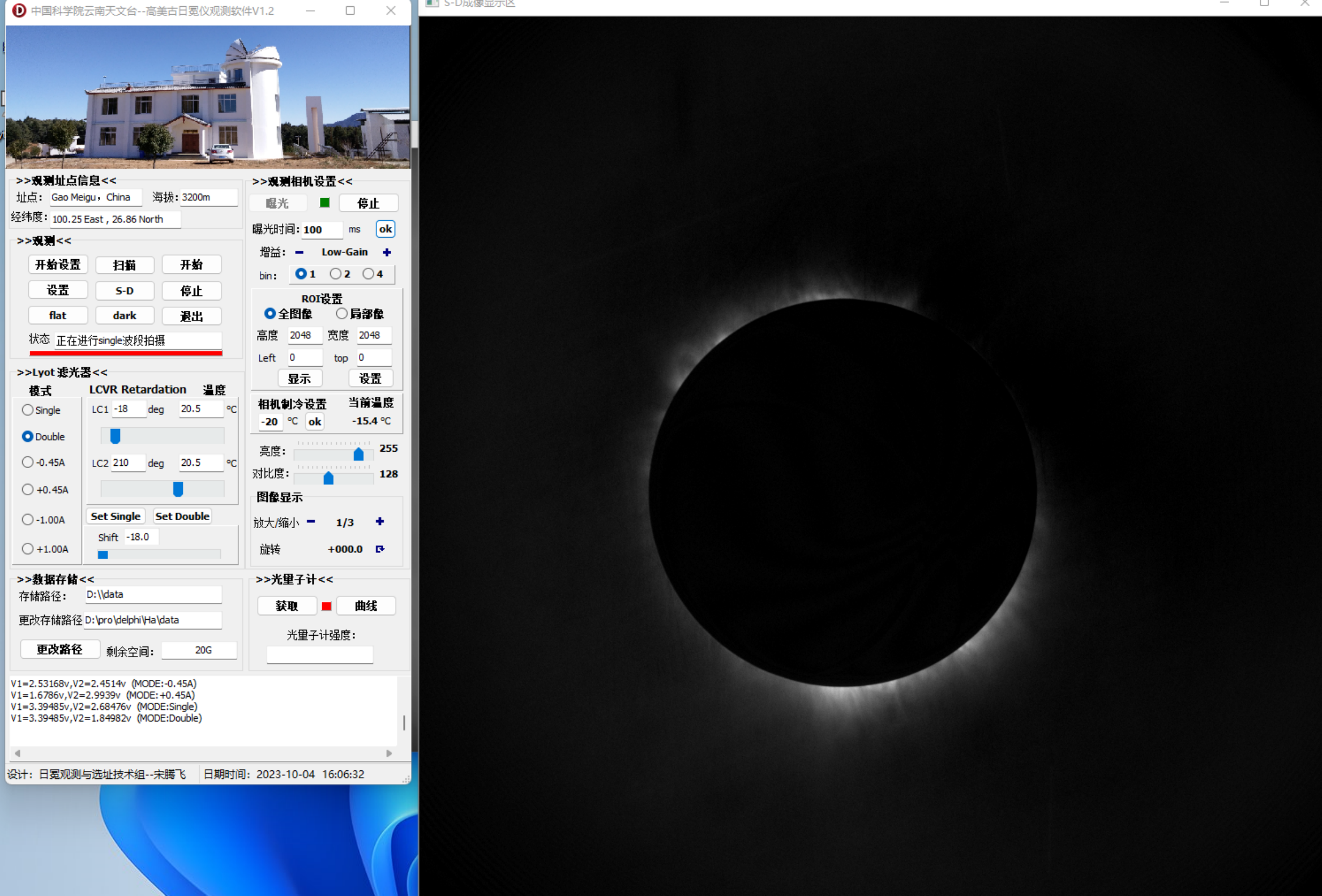}}
\caption{The upgraded operation control system not only enables the control of the coronagraph data acquisition terminal but also adds controls for the equatorial mount's pointing and tracking functions. Additionally, it incorporates real-time wavelength display for the birefringent filter, coronal image visualization, and~status monitoring for the dome and weather~conditions. The red line indicates the data acquisition progress.}
\label{fig5}
\end{figure}

Figure~\ref{fig5} illustrates the user interface of the coronagraph operation control system before and after the upgrade. This intelligent control system demonstrated exceptional operational performance during four years of field observation validation from 2021 to 2024: the overall system availability reached 92\%, with~the proportion of effective operational time far exceeding the industry average; the Mean Time Between Failures (MTBF) reached approximately 1100 h, significantly reducing equipment maintenance frequency and costs; more critically, the~observation time utilization rate increased from 65\% to 90\%, completely resolving the past issue of wasted observation windows caused by delays in manual~operations.

\vspace{-6pt}
\unskip

\subsection{Construction of a Multi-Channel Observation~System}
\label{sec:multi_channel_obs_sys}

The initial YOGIS system was equipped with a 1024 $\times$ 1024 pixel CCD detector, paired with a traditional mechanical shutter. Although~this shutter configuration is a classic setup for early astronomical observation equipment, its limitations have become increasingly apparent when combined with the specific requirements of high-altitude coronal observations in Lijiang: the minute mechanical vibrations generated by the shutter blade movement can cause image blurring and loss of detail; metal transmission components are susceptible to low temperatures, leading to jamming, delays, or~even failure to open/close properly. To~this end, we comprehensively replaced it with a high-performance, cooled CMOS detector with 2048 $\times$ 2048 pixels. Compared to traditional CCDs, the~cooled CMOS not only quadruples the pixel count but also significantly enhances the detail capture capability of the observation field of~view.

\begin{figure}[H]

\begin{adjustwidth}{-\extralength}{0cm}
\centering 
\includegraphics[width=15.0cm, height=6.0cm]{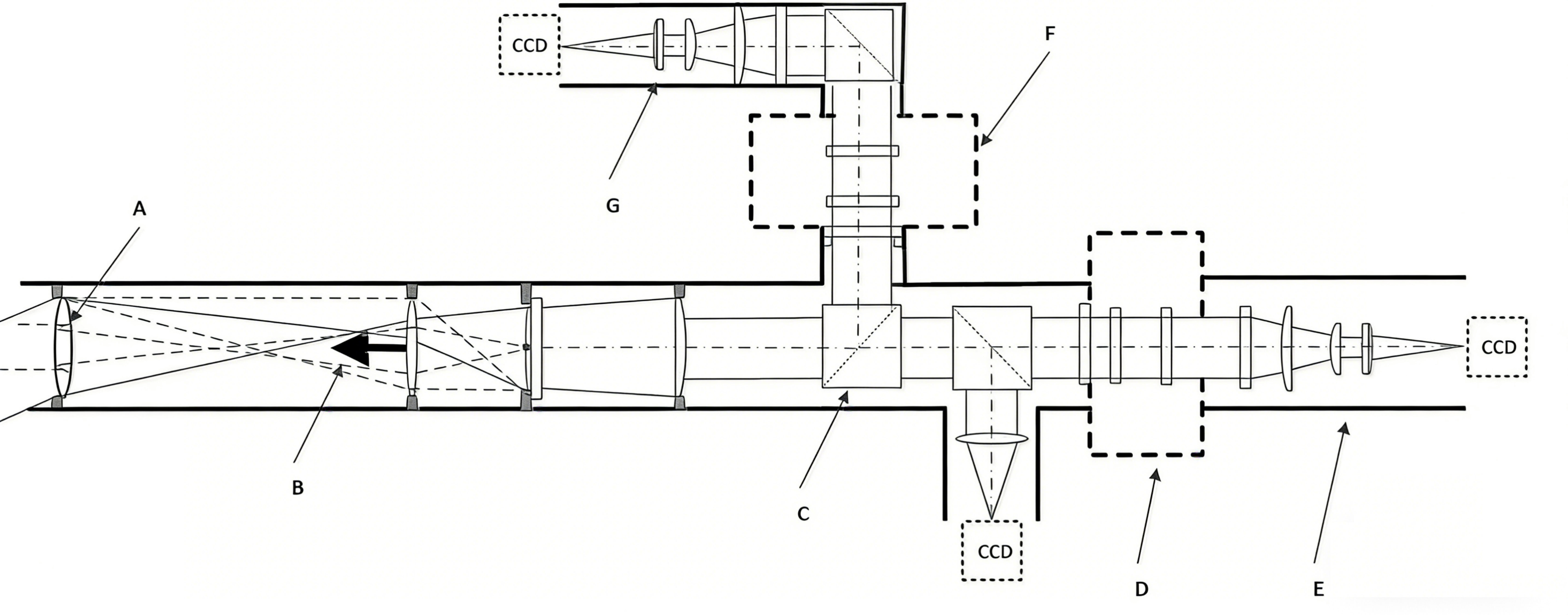}
\end{adjustwidth}

\caption{As shown in the figure, this is the optical design layout of the YOGIS dual-channel observation mode. Here, A: objective lens; B: inner occulter; C: polarizing beam splitter; D: 6374 \AA\ Lyot filter; E: relay lens imaging system; F: 5303 \AA\ Lyot filter; G: relay lens imaging~system.}
\label{fig6}
\end{figure}

To overcome the limitations of single-band observations and expand the dimensions of coronal research, we tackled the core technical challenges of dual-band observations and successfully added a coronal red line observation channel. Figure~\ref{fig6} illustrates the optical design for multi-channel observations with YOGIS, where the red region represents the measurement terminal for the coronal red line. By~employing optical path dispersion compensation technology, we accurately corrected the optical path deviations when switching between the green and red line bands, ensuring the spatial alignment accuracy of dual-band observations. Simultaneously, through collaborative optimization of the detection and control systems, we achieved alternating observations between the two bands. A~typical observation sequence can complete continuous switching as follows: green line single peak $\rightarrow$ red line single peak $\rightarrow$ green line double peak $\rightarrow$ red line double peak, ensuring both observation continuity and avoiding time loss caused by band switching.

\subsection{Breakthroughs in Stray Light Suppression~Technology}
\label{sec:stray_light_suppression}

{Leveraging the dust imaging optical path originally equipped in the NOGIS, we developed a systematic calibration method and established the first quantitative model linking dust distribution to scattering background, enabling effective removal of dust-induced stray light.} 

\begin{figure}[H]

\begin{adjustwidth}{-\extralength}{0cm}
\centering 
\includegraphics[width=15.0cm, height=5.50cm]{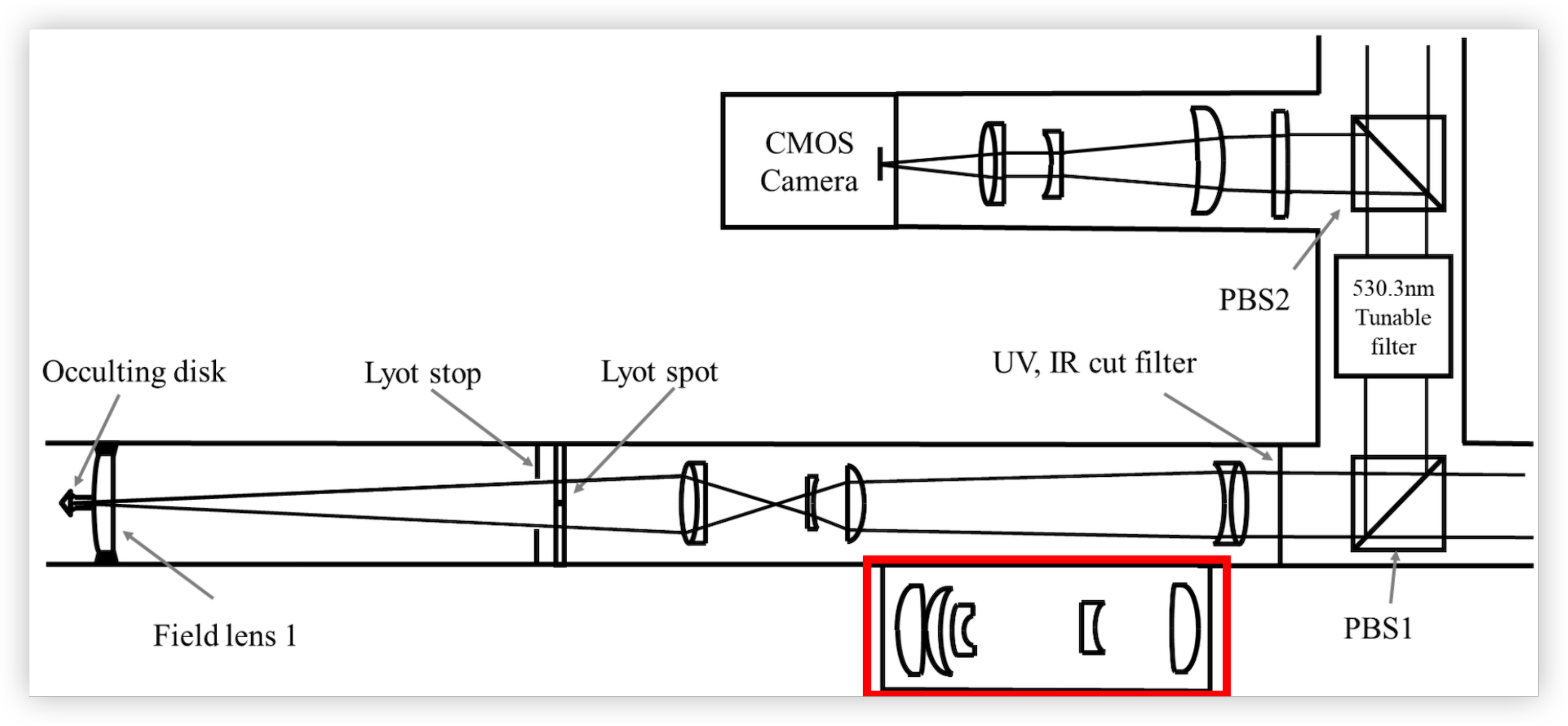}
\end{adjustwidth}

\caption{ Optical layout of the coronagraph in ``dust detection mode'' (the red box indicates the relay lens group for dust imaging). The~system realizes reverse observation of the objective lens surface by reusing the original coronal observation optical path, enabling visualization of dust particles on the objective~lens.}
\label{fig8}
\end{figure}

This design represents an essentially reverse and creative application of the objective lens--imaging system paradigm while fully preserving the original stray light suppression structures within the instrument, such as the non-reflective internal coatings and multi-stage light-blocking apertures. It achieves dust detection while effectively avoiding the strong light interference caused by direct solar illumination, allowing for the inspection of the objective lens' cleanliness without the need for instrument disassembly. Through this specialized optical path modification, we successfully captured panoramic images of the dust distribution on the YOGIS objective lens surface, accurately identifying high-scattering risk areas. This lays a solid physical foundation for the precise correction and calibration of stray light in subsequent coronal observation data. Figure~\ref{fig8} shows the optical layout of the coronagraph in dust detection~mode.

\begin{figure}[H]

\begin{adjustwidth}{-\extralength}{0cm}
\centering 
\includegraphics[width=15.0cm, height=5.50cm]{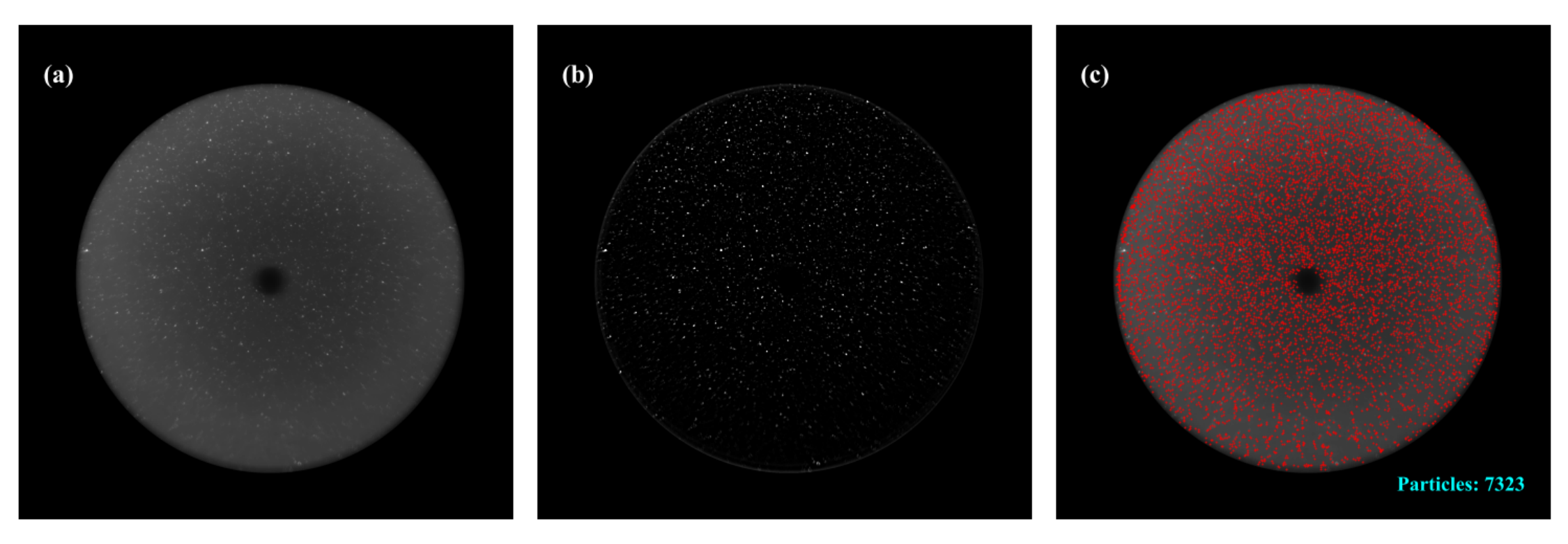}
\end{adjustwidth}

\caption{The captured dust images of the coronagraph objective lens surface, from~left to right, are described as follows: (a) is the dust map of the objective lens surface , the small black dot corresponds to the ``inner occulter disk'' (used for blocking intense light), while the tiny bright spots in the background represent dust particles adhering to the objective lens surface. (b) shows all potential dust signals separated from the original map , dust particles with brightness exceeding the threshold are converted to white bright spots, and~the background is converted to black (via binarization processing with a threshold of 2$\sigma$). (c) presents real dust particles filtered out (with noise excluded) via algorithms based on the binarized result: red spots indicate the identified dust particles, which are used for position calibration and brightness~identification.}
\label{fig9}
\end{figure}

Building upon this optical path detection capability,  we conducted systematic research on the quantitative subtraction of dust-scattering backgrounds and proposed an innovative and practical image differencing method before and after cleaning \citep{Sha2023SoPh}. This method uses the observation image of the objective lens in a clean state as a reference and performs a pixel-level differencing operation with the image obtained when dust is present. This process accurately isolates the pure background signal from dust scattering, leading to the first establishment of a radial attenuation model for the dust-scattering background in coronagraphs. Analysis and validation using YOGIS observation data revealed significant and regular characteristics in dust scattering intensity: the scattering intensity exhibits a strictly linear attenuation trend with increasing heliocentric distance (in solar radii). Furthermore, the~attenuation coefficient is positively correlated with the total dust intensity on the objective lens surface---the greater the amount of dust, the~higher the attenuation coefficient, and~the broader the range of scattering influence. As~shown in Figure~\ref{fig9}, this is the captured dust image of the coronagraph's objective~lens.

Based on this core discovery, we further introduced feature point-matching technology. By~extracting the gray-scale distribution and spatial location features of the dust scattering background, we constructed an adaptive background subtraction algorithm, enabling the precise removal of dust-scattering noise from the original coronal images. This technical solution has been thoroughly validated through actual YOGIS observations.  {After applying the correction, the~mean background intensity decreased significantly, leading to a 30\% improvement in the signal-to-background fluctuation ratio, defined as the ratio of the mean coronal signal to the standard deviation of a background region. This improvement enhances the visibility of faint coronal structures.} This provides reliable data support for subsequent scientific analyses, including the intensity calibration of coronal green and red lines, Doppler velocity measurements, and~the dynamic evolution analysis of coronal plasma. Concurrently, this comprehensive solution, encompassing both optical path design and algorithmic modeling, offers a novel technical approach for ground-based internally-occulted coronagraphs worldwide to address the challenge of dust~scattering.

\vspace{-3pt}
\unskip

\subsection{High-Precision Image Calibration and Multi-Band Registration~Technology}
\label{sec:high_precision_calibration_registration}

\begin{figure}[H]

\begin{adjustwidth}{-\extralength}{0cm}
\centering 
\subfloat{\includegraphics[width=8.0cm, height=6.0cm]{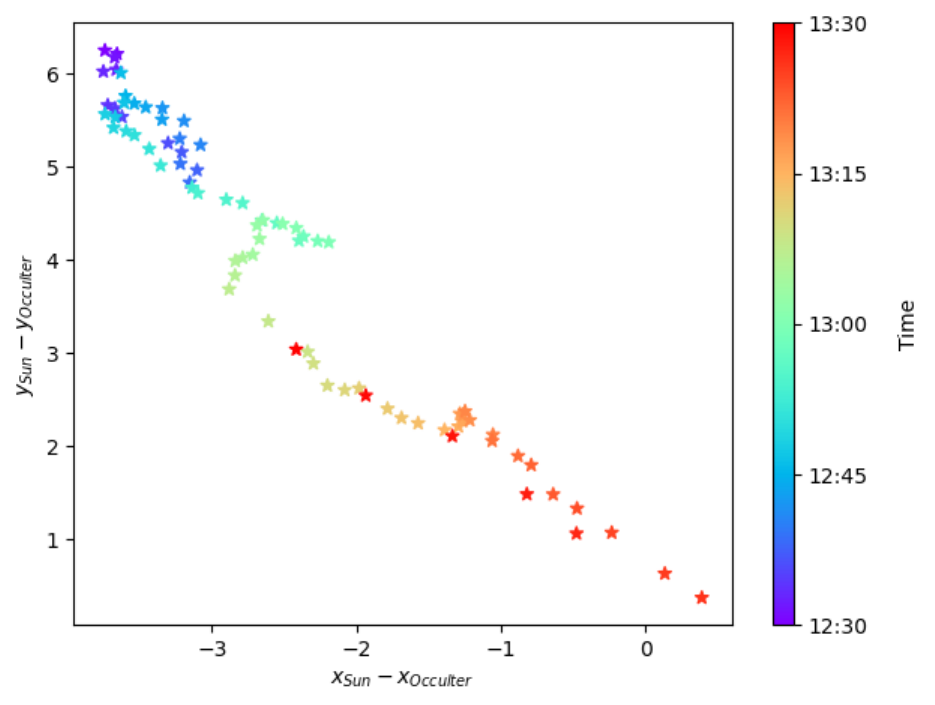}}  
\subfloat{\includegraphics[width=8.0cm, height=6.0cm]{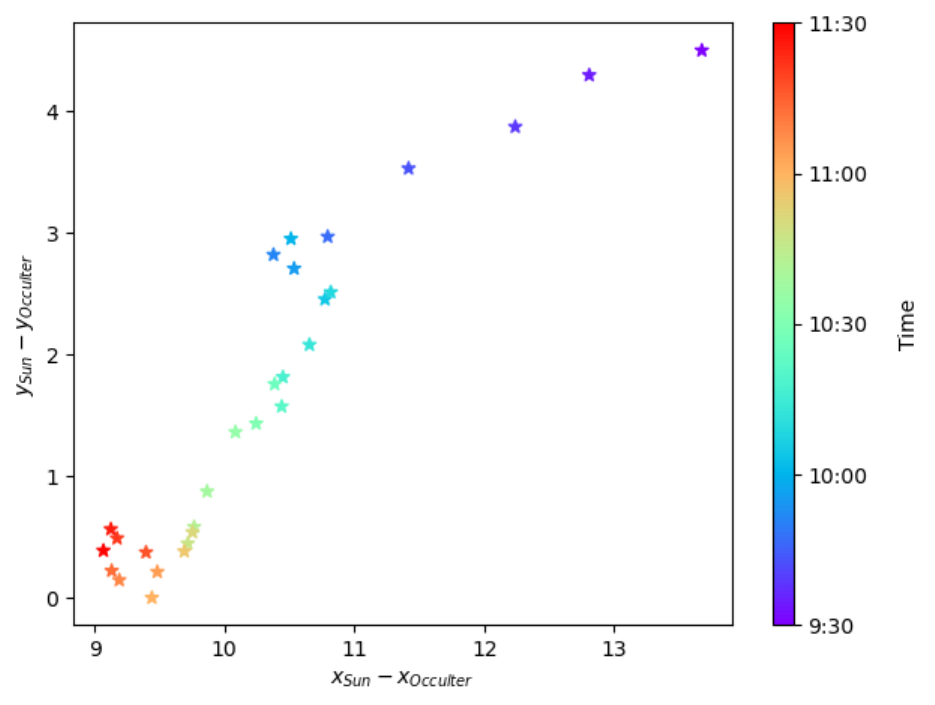}}
\end{adjustwidth}

\caption{ The images show the drift of the solar center relative to the occulter center over observation time during coronal observations (observation date: {11 November 2024}; unit: pixel). The~(\textbf{left}) and (\textbf{right}) panels illustrate the horizontal (x-axis) and vertical (y-axis) drift of YOGIS and SICG, respectively. Different colors correspond to the observation times of the day, with~relatively small drift values near~noon.}
\label{fig10}
\end{figure}

Ground-based coronagraphs observe the low corona by creating an artificial solar eclipse using an occulter disk. However, due to factors such as pointing accuracy and temperature fluctuations, the~occulter disk cannot always be precisely aligned with the Sun. Figure~\ref{fig10} shows the variation in the solar center offset with time during a day's observation. Furthermore, the~observed images lack clear landmarks such as the solar limb, making it difficult to accurately determine key parameters like helioprojective coordinates and the solar radius. This severely limits the joint analysis of multi-instrument and multi-band~data.

\vspace{-9pt}

The core approach to solving this problem is to use calibrated space-based images with known heliocentric coordinates, solar radius, and~north--south pole orientation as a reference. By~performing operations such as translation, rotation, and~scaling, the~ground-based coronagraph images are registered with these reference images, thereby indirectly obtaining the aforementioned key~parameters.

Sha Feiyang~et~al. proposed an Automated High-Precision Registration Algorithm (APRIL). Using SDO/AIA 211 \AA\ extreme ultraviolet (EUV) images as the reference, APRIL achieves precise registration between ground-based coronal green line images and the reference images \citep{sha2025ApJ}. This enables the successful mapping of coronal images onto the Helioprojective Cartesian Coordinate (HCC) system. Under~optimal data quality conditions, the~registration accuracy is better than 0.1$^\prime$, and~in most cases, it is no worse than 0.4$^\prime$.

This method has been thoroughly validated using observational data from two coronagraphs: the YOGIS and SICG. The~validation covers 100 days of data from 2013 to 2024, as~well as 34 days of statistical data from 2015 and 2024. These tests confirm its broad applicability, effectively supporting studies of coronal transient activities and joint data analysis across multiple instruments. Additionally, it provides technical support for improving the pointing accuracy of coronagraphs. It should be noted that this method assumes the presence of clear coronal structures around the solar disk and is not suitable for coronal images during solar minimum~periods.

\section{Data~Applications}
\label{sec:data_applications}
\unskip

\subsection{Coronal Structure and Activity~Observations}
\label{subsec:coronal_structure_observations}

Using the 2D coronal green-line images from the YOGIS, combined with photospheric magnetograms from the SDO/HMI instrument, Zhang~et~al. \citep{Zhang2022RAA007Z} calculated the coronal magnetic field strength via the Potential Field Source Surface (PFSS) model and conducted an in-depth correlation analysis between the two datasets. Their study revealed that the correlation coefficient between the green-line intensity and magnetic field strength reaches a maximum of 0.82 at a height of 1.1~$R_{\odot}$, which corresponds exactly to the apex of closed coronal loops. In~terms of loop distribution, the~correlation coefficient at the loop apex is consistently above 0.8, while the correlation at the loop footpoints is weaker and exhibits significant fluctuations. Additionally, in~extended studies covering the $\pm40^\circ$ latitude range and a complete Carrington rotation (CR 2143, 27 days), the~maximum correlation coefficient consistently appears at 1.1~$R_{\odot}$. This finding clarifies the special role of the 1.1~$R_{\odot}$ height in coronal physics research: {It not only yields key insights into the origin of the slow solar wind (implying this height may represent a critical source region for the slow solar wind) but also paves new ways for coronal heating mechanism research, as~Squire~et~al. have shown that solar wind acceleration and heating commence in the lower corona} \citep{Squire2022NatAs}. {These results can be further validated against state-of-the-art MHD simulations of the solar corona, such as the eclipse prediction studies by Linker~et~al.} \citep{Linker2024AGUF, Linker2019AGUF}.

To build a bridge for multi-wavelength coronal research, Zhang~et~al. \citep{Zhang2022RAA012Z} conducted comparative analyses between green-line data from theYOGIS and EUV band data from the SDO/AIA instrument. They investigated the correlation coefficient between the coronal green line and the 211~\AA\ from 0.89 to 0.99.  The high correlation coefficient between the coronal green line and the 211~\AA\ not only intuitively reflects the physical essence of ``unified observation targets'' but also carries profound scientific significance and application value. {Furthermore, the~unique scientific merit of YOGIS lies in its capability to measure the line-of-sight velocity of coronal plasma---a parameter not available from AIA images. This velocity information, derived from the $\pm 0.45$ \AA~line-wing images obtained with the fast-tunable Lyot filter, enables studies of coronal dynamics, wave propagation, and~mass flows that complement the structural information provided by space-based EUV observations.}  {This result confirms the consistency of the physical state of the inner coronal plasma, forms a complementary connection between ground-based and space-based observations, and~offers a new approach for addressing core scientific issues such as coronal heating and magnetic field diagnostics. This finding also provides guidance for future research: through joint observations of these two bands, the~physical processes (e.g., slow solar wind acceleration) at the critical 1.1 $R_{\odot}$ height can be further explored, which can support the design of observational schemes for next-generation coronagraphs.}

\subsection{Early Warning of~CMEs}
\label{subsec:Early Warning of CMEs}

\begin{figure}[H]

\begin{adjustwidth}{-\extralength}{0cm}
\centering 
\subfloat{\includegraphics[width=8.0cm, height=6.0cm]{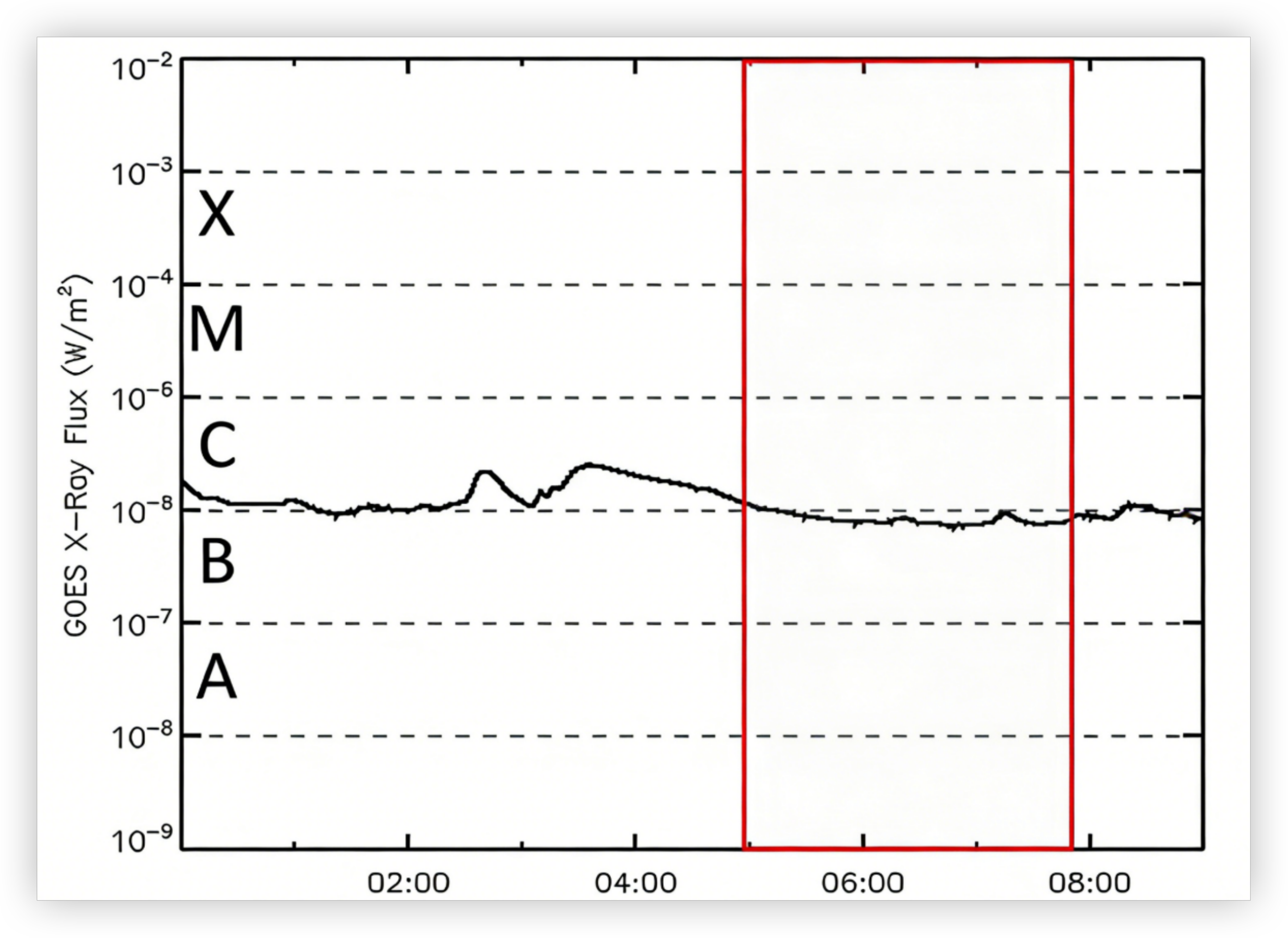}}  
\subfloat{\includegraphics[width=6.0cm, height=6.0cm]{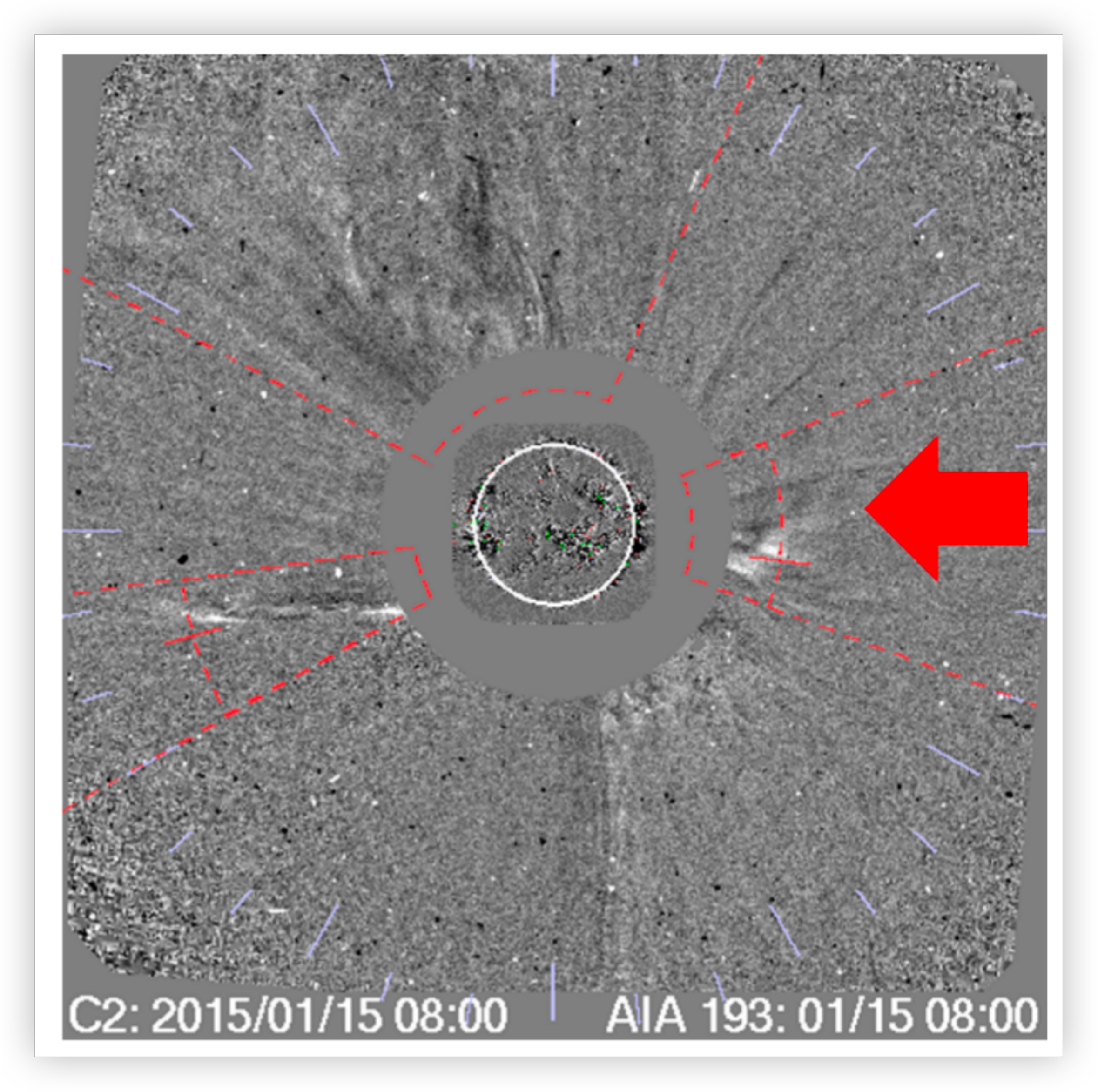}}
\end{adjustwidth}

\caption{\textbf{Left:} GOES soft X-ray flux showing a C-class flare on 15 January 2015. The red box highlights the time interval (05:00---06:30 UT) when the bright structure was observed by YOGIS. \textbf{Right:} SOHO/LASCO C2 image at 08:00 UT, with the red dashed line indicating the position angle of the CME. The close alignment between the YOGIS detection and the LASCO CME confirms the early warning capability of ground-based coronagraphs.}
\label{fig11}
\end{figure}

\begin{figure}[H]
\includegraphics[width=12.0cm, height=5.50cm]{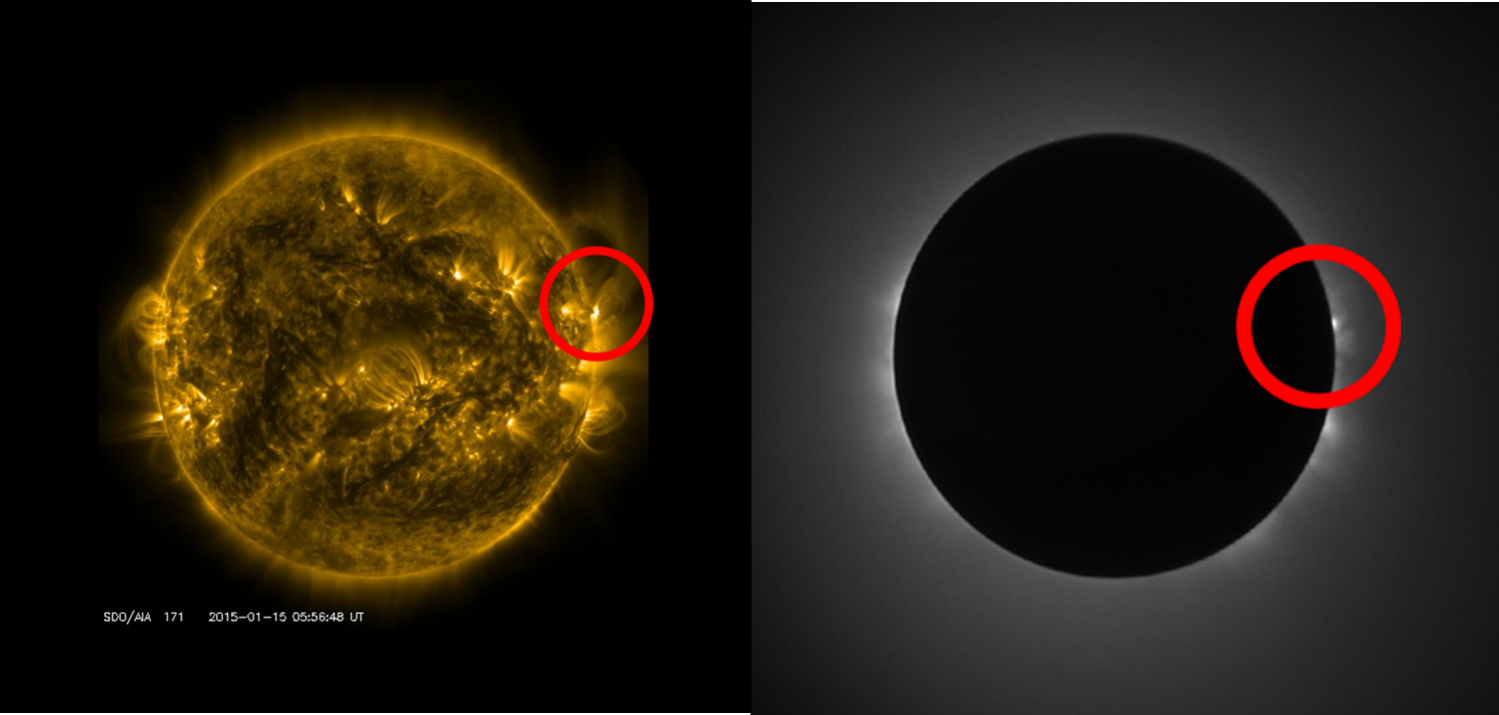}

\caption{ {Global comparison of coronal observations on {15 January 2015}. (\textbf{Left}): SDO/AIA 171~\AA~image showing the full solar disk, highlighting the active-region coronal loops at the western limb (red circle); (\textbf{right}): corresponding YOGIS coronal image, with~the occulting disk blocking the bright solar disk, revealing the faint coronal structures at the same limb location (red circle).}}
\label{fig12}
\end{figure}

Regarding CMEs, the high time-resolution green-line observations of the YOGIS in fast mode (15 s per frame) can accurately capture their dynamic evolutionary processes. Leveraging its high time-resolution observational capability, the~instrument can promptly detect relevant propagation signals when a CME just breaks through the solar surface (at approximately 1.03 $R_{\odot}$). Figure~\ref{fig11} shows that {following a C-class flare, a~bright structure was observed on the western limb by YOGIS. Approximately one hour later, the~CME was detected by the Large Angle and Spectrometric Coronagraph (LASCO) aboard the Solar and Heliospheric Observatory (SOHO) satellite at a similar but not identical position angle, consistent with the typical expansion and deflection of CMEs during their initial propagation phase} \citep{2024JGRA}. The~brightness of this bright structure began to increase at 05:00 (UT) and ceased at 06:30; during the late evolutionary stage, a~cusp-shaped structure was clearly observed, with~its apex continuously moving outward away from the solar limb (Figure~\ref{fig12}). Calculations indicate that the average velocity of the brightest point of the bright structure in the sky plane was approximately 20 $km {s}^{-1}$. Near~the brightness peak, the~velocity varied from 25 to 15 $km {s}^{-1}$. This observational achievement provides a valuable case study for the early identification of CMEs and the calculation of their dynamical parameters, fully demonstrating the crucial role of the YOGIS in the early warning of~CMEs.

\vspace{-12pt}
\vspace{-6pt}

\unskip

\section{Future Plans for~YOGIS}
\label{sec:Future_YOGIS}

Despite multiple technical upgrades and significant scientific achievements made with YOGIS, the~analysis of fine-scale coronal structures remains severely constrained, limiting its further development and application.
Overcoming this limitation requires breakthroughs in the fabrication of large-aperture objective lenses.
{Building a 500 mm coronagraph is a natural and necessary next step, and~the technical experience accumulated with YOGIS---particularly in automation, light scattering correction, and~dust \mbox{handling---has} well prepared our team for this endeavor.}
{Furthermore, } the coronal green line, owing to its high intensity, strong magnetic sensitivity, and~its diagnostic potential for high-temperature coronal plasma, makes polarimetric measurement a crucial tool for directly probing coronal magnetic fields.
{We} plan to upgrade the acquisition terminal of YOGIS by inserting a polarization modulation module---consisting of a rotating waveplate and a static analyzer---downstream of the Lyot birefringent filter. This will enable routine polarimetric measurements, allowing us to study the spatiotemporal evolution of the green line's polarization degree and angle. Such measurements can trace the velocity of coronal plasma flows and detect the onset and propagation of CMEs. Moreover, time-series analysis of the polarization signals will permit the capture of dynamic processes in fine-scale structures such as coronal plumes and~jets.

Since its commissioning in 2013, YOGIS has operated under a paradigm of fixed-interval acquisition followed by post-processing. 
 The~inherent limitations of this scheduled acquisition mode have long remained unresolved. The~non-stationary characteristics of stray light, which gradually evolve with the slow degradation of mirror surface contamination, cause systematic day-to-day variations in the signal-to-noise ratio of coronal images. Under~the fixed-interval paradigm, data quality at the moment of acquisition is inherently unpredictable. Moreover, the~inability to precisely locate the spatial origin of coronal transient events hinders reliable linkage to their source regions on the solar disk. Although~coordinate deviations can be partially corrected in post-processing via manual alignment of morphological features, this approach has become an insurmountable bottleneck for tasks requiring real-time transient monitoring and operational space weather forecasting. Drawing on more than a decade of experience in operating and maintaining a high-altitude coronagraph, the~ultimate goal for YOGIS is to achieve unattended, routine coronal observations. This intelligent and autonomous vision comprises four key~capabilities:
 \begin{itemize}
    \item Autonomous judgment of observable windows based on real‑time measurements of cloud cover and sky background brightness.
    \item Real‑time monitoring of the operating status of critical components, with~automatic failover to backup systems or execution of safety procedures upon anomaly detection.
    \item Event‑driven mode switching triggered when specific physical parameters (e.g., total intensity, degree of polarization) extracted from coronal images exceed preset~\mbox{thresholds}.
    \item Automated generation and dissemination of data products to scientific databases.
    \end{itemize}

\begin{figure}[H]

\begin{adjustwidth}{-\extralength}{0cm}
\centering 
\includegraphics[width=15.0cm, height=5.50cm]{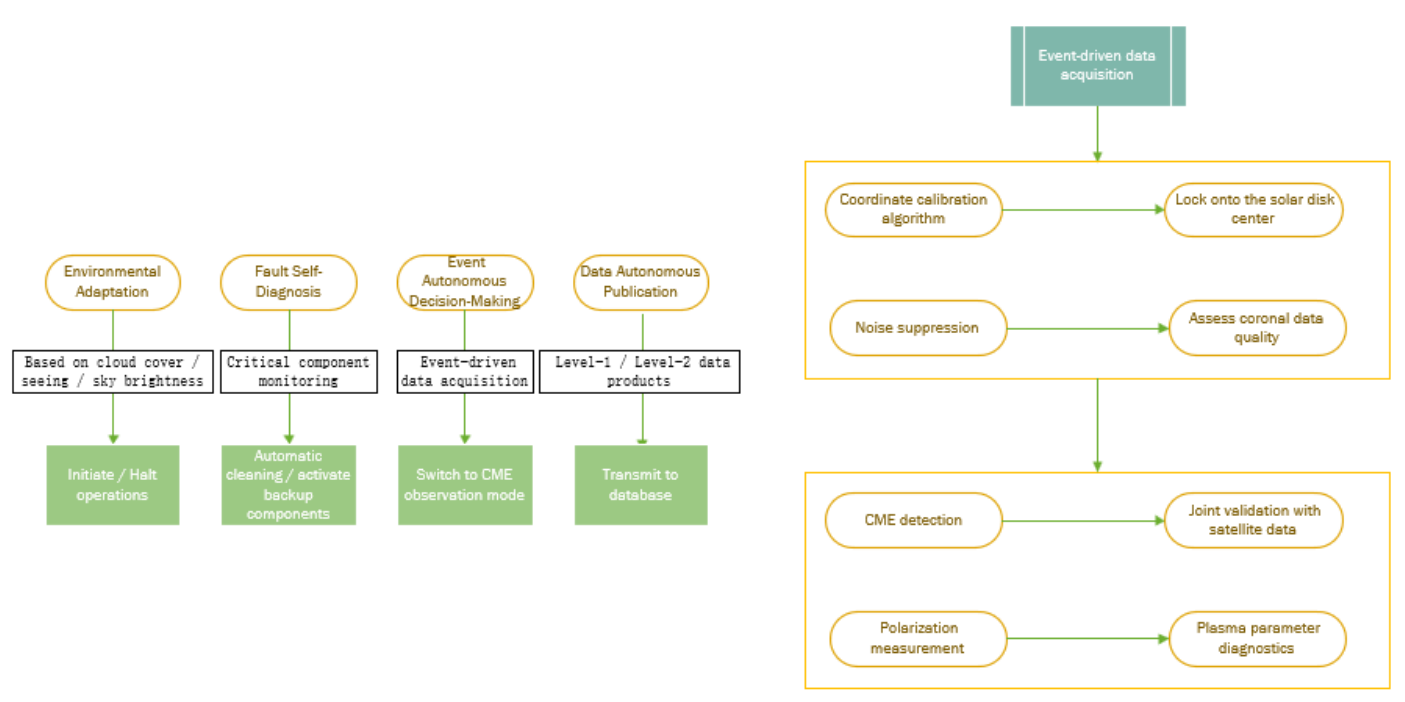}
\end{adjustwidth}

\caption{This figure depicts the complete technical architecture of the future fully autonomous observing mode of the YOGIS coronagraph. (\textbf{Left}): the end-to-end workflow for unattended operation, encompassing autonomous observation condition assessment, equipment status management, event-triggered acquisition of CMEs, and~automated data publication. (\textbf{Right}): the detection pipeline for CME eruptions, ranging from identification of the solar disk center coordinates and coronal image noise suppression to polarimetric measurements and physical parameter inversion. The~two modules operate in synergy, enabling a paradigm shift from “scheduled acquisition, post-processing” to “event-driven, real-time response, intelligent autonomous” observations.}
\label{fig13}
\end{figure}

Realizing such fully intelligent autonomous observations is also a core requirement for establishing coronagraph stations at ultra‑high-altitude sites. Higher altitudes offer thinner atmospheres and substantially lower sky-scattered light, providing superior conditions for ground‑based coronal diagnostics. However, these advantages come at the cost of severe hypoxia, difficult terrain, and~limited accessibility, imposing extreme challenges on observer safety, equipment transportation, and~long‑term operational expenditure. In~the landscape of ground‑based coronagraphy, YOGIS is poised to evolve into a new generation of rapid‑response, fully automatic coronagraph. Figure~\ref{fig13} illustrates the envisioned fully autonomous observing mode of~YOGIS.

YOGIS requires not only technological upgrades but also sustained advancements in scientific frontiers and application extensions. Single-platform, single-wavelength observations are inherently insufficient to simultaneously disentangle the complex coupling among the coronal magnetic field, velocity field, and~temperature field. There is an urgent need to develop a synergistic ground-based and space-borne, multi-perspective, multi-wavelength joint detection system. In~this context, collaborative observations integrating international forefront solar facilities---such as the Daniel K. Inouye Solar Telescope (DKIST), the~SDO/AIA, and~the Upgraded Coronal Multi-channel Polarimeter (UCoMP)---are imperative. {We also plan to establish an open‑access database for YOGIS data to foster global collaboration. With~China's  growing capabilities in solar observations, we aim to support the international community by sharing our expertise and data, contributing to a worldwide ground‑based coronal observation network. Ground‑based coronagraphs remain essential globally, complementing space missions and providing long‑term monitoring that benefits all.}

Addressing the challenge of coronal magnetometry through synergistic observations, constructing an integrated space–ground detection network with homegrown instruments, empowering precise CME forecasting with artificial intelligence, and~leading international cooperation via data sharing—these four mutually reinforcing and progressively advancing directions collectively define the contemporary frontier of coronal physics and space weather~science.

\vspace{6pt} 

\authorcontributions{Conceptualization, X.Z. and Y.L.; methodology, X.Z., T.S., and~M.S.; software, M.Z.; investigation, X.L. (Xiaobo Li); data curation, F.S. and X.L. (Xiande Liu); writing---original draft preparation, X.Z.; writing---review and editing, X.Z. and Y.L.; funding acquisition, Y.L. All authors have read and agreed to the published version of the manuscript.}

\funding{This work was jointly supported by multiple funding sources. Specifically, it received partial support from the National Natural Science Foundation of China (NSFC grant Nos. 12173086, 12373063, 11533009, 12163004, 12473089, and~42274227). Additionally, the~Yunnan Fundamental Research Projects (grant Nos. 202501AS070004 and 202401AT070140), and~the Yunnan Key Laboratory of Solar Physics and Space Science (grant No. 202205AG070009).}

\dataavailability{Data are contained within this article.} 

\acknowledgments{{We sincerely thank the NAOJ team---particularly Takashi Sakurai, Kiyoshi Ichimoto, Masaoki Hagino, Goichi kimura, Motokazu Noguchi, Nobuyuki Tanaka, Kazuyoshi Kumagai, Kazuya Shinoda, Tetsuo Nishino, Takeo Fukuda, and~their colleagues---for their pioneering development of the original NOGIS coronagraph. Their foundational work has made the YOGIS instrument possible, and~we are grateful for their long-standing collaboration and~support.} We also acknowledge the data resources from the National Space Science Data Center, National Science and Technology Infrastructure of China (\url{http://www.nssdc.ac.cn}). We acknowledge the Chinese Meridian Project for providing high-quality data from the SICG. We also thank the NASA/SDO and the AIA science team for open data access. We thank the anonymous referees for helpful comments and suggestions on this~manuscript.}

\conflictsofinterest{The authors declare no conflicts of~interest.} 

\begin{adjustwidth}{-\extralength}{0cm}
\reftitle{References}

\PublishersNote{}

\end{adjustwidth}


\begin{thebibliography}{999}

\bibitem[{Wang} et~al.(1997){Wang}, {Sheeley}, {Hawley}, {Kraemer},
  {Brueckner}, {Howard}, {Korendyke}, {Michels}, {Moulton}, {Socker}, and
  {Schwenn}]{wang1997ApJ}
{Wang}, Y.M.; {Sheeley}, N.R., Jr.; {Hawley}, S.H.; {Kraemer}, J.R.;
  {Brueckner}, G.E.; {Howard}, R.A.; {Korendyke}, C.M.; {Michels}, D.J.;
  {Moulton}, N.E.; {Socker}, D.G.;  et~al.
\newblock {The Green Line Corona and Its Relation to the Photospheric Magnetic
  Field}.
\newblock {\em \apj} {\bf 1997}, {\em 485},~419--429.
\newblock {\url{https://doi.org/10.1086/304405}}.

\bibitem[{Yang} et~al.(2024){Yang}, {Tian}, {Tomczyk}, {Liu}, {Gibson},
  {Morton}, and {Downs}]{yang2024Sci}
{Yang}, Z.; {Tian}, H.; {Tomczyk}, S.; {Liu}, X.; {Gibson}, S.; {Morton}, R.J.;
  {Downs}, C.
\newblock {Observing the evolution of the Sun's global coronal magnetic field
  over 8 months}.
\newblock {\em Science} {\bf 2024}, {\em 386},~76--82.
\newblock {\url{https://doi.org/10.1126/science.ado2993}}.

\bibitem[{Chen} et~al.(2023){Chen}, {Bai}, {Tian}, {Li}, {Chen}, {Yang}, and
  {Yang}]{Chen2023MNRAS}
{Chen}, Y.; {Bai}, X.; {Tian}, H.; {Li}, W.; {Chen}, F.; {Yang}, Z.; {Yang}, Y.
\newblock {Solar coronal magnetic field measurements using spectral lines
  available in Hinode/EIS observations: Strong and weak field techniques and
  temperature diagnostics}.
\newblock {\em \mnras} {\bf 2023}, {\em 521},~1479--1488.
\newblock {\url{https://doi.org/10.1093/mnras/stad583}}.

\bibitem[{Song} et~al.(2025){Song}, {Wang}, {Li}, {Wang}, and
  {Chen}]{song2025ApJ}
{Song}, H.; {Wang}, R.; {Li}, L.; {Wang}, B.; {Chen}, Y.
\newblock {On the Nature of the Bright Front of Solar Coronal Mass Ejections}.
\newblock {\em \apj} {\bf 2025}, {\em 988},~270--279.
\newblock {\url{https://doi.org/10.3847/1538-4357/adec88}}.

\bibitem[{Lyot}(1939)]{Lyot1939MNRAS}
{Lyot}, B.
\newblock {The study of the solar corona and prominences without eclipses
  (George Darwin Lecture, 1939)}.
\newblock {\em \mnras} {\bf 1939}, {\em 99},~580--596.
\newblock {\url{https://doi.org/10.1093/mnras/99.8.580}}.

\bibitem[{Tian} et~al.(2013){Tian}, {Tomczyk}, {McIntosh}, {Bethge}, {de Toma},
  and {Gibson}]{Tian2013SoPh}
{Tian}, H.; {Tomczyk}, S.; {McIntosh}, S.W.; {Bethge}, C.; {de Toma}, G.;
  {Gibson}, S.
\newblock {Observations of Coronal Mass Ejections with the Coronal Multichannel
  Polarimeter}.
\newblock {\em \solphys} {\bf 2013}, {\em 288},~637--650.
\newblock {\url{https://doi.org/10.1007/s11207-013-0317-5}}.

\bibitem[{de Wijn} et~al.(2012){de Wijn}, {Burkepile}, {Tomczyk}, {Nelson},
  {Huang}, and {Gallagher}]{Wijn2012SPIE}
{de Wijn}, A.G.; {Burkepile}, J.T.; {Tomczyk}, S.; {Nelson}, P.G.; {Huang}, P.;
  {Gallagher}, D.
\newblock {Stray light and polarimetry considerations for the COSMO
  K-Coronagraph}.
\newblock In \emph{Ground-Based and Airborne Telescopes IV, Proceedings of the  SPIE Astronomical Telescopes + Instrumentation, Amsterdam, The Netherlands, 1--6 July 2012}
;
  {Stepp}, L.M., {Gilmozzi}, R., {Hall}, H.J., Eds.;  {Society of Photo-Optical Instrumentation Engineers (SPIE)
  Conference Series}; SPIE: Bellingham, WA, USA, 2012; Volume  8444, p. 84443N.
\newblock {\url{https://doi.org/10.1117/12.926511}}.

\bibitem[{Liang} et~al.(2021){Liang}, {Qu}, {Chen}, {Zhong}, {Song}, and
  {Li}]{Liang2021MNRAS}
{Liang}, Y.; {Qu}, Z.Q.; {Chen}, Y.J.; {Zhong}, Y.; {Song}, Z.M.; {Li}, S.Y.
\newblock {Registration and imaging polarimetry of the Fe 6374 {\r{A}} red
  coronal line during the 2017 total solar eclipse}.
\newblock {\em \mnras} {\bf 2021}, {\em 503},~5715--5729.
\newblock {\url{https://doi.org/10.1093/mnras/stab463}}.

\bibitem[{Han} et~al.(2021){Han}, {Yang}, {Liu}, {Zhang}, {Qing}, {Li}, {Wu},
  and {Luo}]{han2021MNRAS}
{Han}, Y.; {Yang}, Q.; {Liu}, N.; {Zhang}, K.; {Qing}, C.; {Li}, X.; {Wu}, X.;
  {Luo}, T.
\newblock {Analysis of wind-speed profiles and optical turbulence above
  Gaomeigu and the Tibetan Plateau using ERA5 data}.
\newblock {\em \mnras} {\bf 2021}, {\em 501},~4692--4702.
\newblock {\url{https://doi.org/10.1093/mnras/staa2960}}.

\bibitem[{Rao} et~al.(2016){Rao}, {Zhu}, {Rao}, {Zhang}, {Bao}, {Kong}, {Guo},
  {Zhong}, {Ma}, {Li}, {Wang}, {Zhang}, {Fan}, {Chen}, {Feng}, {Gu}, and
  {Liu}]{rao2016ApJ}
{Rao}, C.; {Zhu}, L.; {Rao}, X.; {Zhang}, L.; {Bao}, H.; {Kong}, L.; {Guo}, Y.;
  {Zhong}, L.; {Ma}, X.; {Li}, M.;  et~al.
\newblock {Instrument Description and Performance Evaluation of a High-Order
  Adaptive Optics System for the 1 m New Vacuum Solar Telescope at Fuxian Solar
  Observatory}.
\newblock {\em \apj} {\bf 2016}, {\em 833},~210--219.
\newblock {\url{https://doi.org/10.3847/1538-4357/833/2/210}}.

\bibitem[{Wu} et~al.(2014){Wu}, {Qian}, {Huang}, {Wang}, {Cui}, and
  {Qing}]{Wu2014AcASn}
{Wu}, X.Q.; {Qian}, X.M.; {Huang}, H.H.; {Wang}, P.; {Cui}, C.L.; {Qing}, C.
\newblock {Measurements of Seeing, Isoplanatic Angle, and Coherence Time by
  Using Balloon-borne Microthermal Probes at Gaomeigu}.
\newblock {\em Acta Astron. Sin.} {\bf 2014}, {\em 55},~144--153.

\bibitem[{Ichimoto} et~al.(1999){Ichimoto}, {Noguchi}, {Tanaka}, {Kumagai},
  {Shinoda}, {Nishino}, {Fukuda}, {Sakurai}, and {Takeyama}]{ichinoto1999PASJ}
{Ichimoto}, K.; {Noguchi}, M.; {Tanaka}, N.; {Kumagai}, K.; {Shinoda}, K.;
  {Nishino}, T.; {Fukuda}, T.; {Sakurai}, T.; {Takeyama}, N.
\newblock {A New Imaging System of the Corona at Norikura}.
\newblock {\em \pasj} {\bf 1999}, {\em 51},~383--391.
\newblock {\url{https://doi.org/10.1093/pasj/51.3.383}}.

\bibitem[{Sakurai}(2012)]{sakurai2012ASPC}
{Sakurai}, T.
\newblock {Sixty Years of Norikura Solar Observatory}.
\newblock In \emph{Proceedings of the Hinode-3: The 3rd Hinode Science Meeting};
  {Sekii}, T., {Watanabe}, T., {Sakurai}, T., Eds.;  
  {Astronomical Society of the Pacific Conference Series}; Astronomical Society of the Pacific: San~Franciso, CA, USA, 2012; Volume 454, p. 439.

\bibitem[{Song} et~al.(2025){Song}, {Liu}, {Zhang}, {Zhao}, {Li}, {Luo}, {Sha},
  {Liu}, {Oloketuyi}, and {Wang}]{song2025Univ}
{Song}, T.; {Liu}, Y.; {Zhang}, X.; {Zhao}, M.; {Li}, X.; {Luo}, Q.; {Sha}, F.;
  {Liu}, Q.; {Oloketuyi}, J.; {Wang}, X.
\newblock {Toward Automated Coronal Observations: A New Integrated System Based
  on the Lijiang 10 cm Coronagraph}.
\newblock {\em Universe} {\bf 2025}, {\em 11},~154--166.
\newblock {\url{https://doi.org/10.3390/universe11050154}}.

\bibitem[{Xu} et~al.(2023){Xu}, {Feng}, {Pu}, {Wang}, {Cao}, {Ren}, {Zhang},
  {Ma}, {Bai}, {Esamdin}, {Li}, {Tian}, {Wang}, {Zhao}, and {Shi}]{Xu2023RAA}
{Xu}, J.; {Feng}, G.-j.; {Pu}, G.-x.; {Wang}, L.-t.; {Cao}, Z.-H.; {Ren}, L.-Q.;
  {Zhang}, X.; {Ma}, S.-g.; {Bai}, C.-h.; {Esamdin}, A.;  et~al.
\newblock {Site-testing at the Muztagh-ata Site V. Nighttime Cloud Amount
  during the Last Five Years}.
\newblock {\em Res. Astron. Astrophys.} {\bf 2023}, {\em
  23},~045015.
\newblock {\url{https://doi.org/10.1088/1674-4527/acc29b}}.

\bibitem[{Lemen} et~al.(2012){Lemen}, {Title}, {Akin}, {Boerner}, {Chou},
  {Drake}, {Duncan}, {Edwards}, {Friedlaender}, {Heyman}, {Hurlburt}, {Katz},
  {Kushner}, {Levay}, {Lindgren}, {Mathur}, {McFeaters}, {Mitchell}, {Rehse},
  {Schrijver}, {Springer}, {Stern}, {Tarbell}, {Wuelser}, {Wolfson}, {Yanari},
  {Bookbinder}, {Cheimets}, {Caldwell}, {Deluca}, {Gates}, {Golub}, {Park},
  {Podgorski}, {Bush}, {Scherrer}, {Gummin}, {Smith}, {Auker}, {Jerram},
  {Pool}, {Soufli}, {Windt}, {Beardsley}, {Clapp}, {Lang}, and
  {Waltham}]{Lemen2012SoPh}
{Lemen}, J.R.; {Title}, A.M.; {Akin}, D.J.; {Boerner}, P.F.; {Chou}, C.;
  {Drake}, J.F.; {Duncan}, D.W.; {Edwards}, C.G.; {Friedlaender}, F.M.;
  {Heyman}, G.F.;  et~al.
\newblock {The Atmospheric Imaging Assembly (AIA) on the Solar Dynamics
  Observatory (SDO)}.
\newblock {\em \solphys} {\bf 2012}, {\em 275},~17--40.
\newblock {\url{https://doi.org/10.1007/s11207-011-9776-8}}.

\bibitem[{Zhang}(2023)]{Zhang2023soma}
{Zhang}, H.
\newblock {\em {Solar Magnetism}};  Springer: New York, NY, USA, 2023.
\newblock {\url{https://doi.org/10.1007/978-981-99-1759-4}}.

\bibitem[{Gong} and {Socker}(2004)]{gong2004SPIE}
{Gong}, Q.; {Socker}, D.
\newblock {Theoretical study of the occulted solar coronagraph}.
\newblock In \emph{Optical Systems Degradation, Contamination, and
  Stray Light: Effects, Measurements, and Control,  Proceedings of the Optical Science and Technology, the SPIE 49th Annual Meeting, Denver, CO, USA, 2--6 August 2004}; {Chen}, P.T.C., {Fleming},
  J.C., {Dittman}, M.G., Eds.;   {Society of
  Photo-Optical Instrumentation Engineers (SPIE) Conference Series}; SPIE: Bellingham, WA, USA, 2004; Volume 5526,  pp.
  208--219.
\newblock {\url{https://doi.org/10.1117/12.549275}}.

\clearpage
\bibitem[{Liu} et~al.(2021){Liu}, {Zhang}, {Song}, {Sun}, {Liu}, {Wang},
  {Zhao}, {Zhang}, {Xu}, {Fu}, {Pi}, {Huang}, {Li}, {Fu}, {Fan}, {Liu}, {Shen},
  {Sha}, {Li}, {Jin}, {Liu}, {Xia}, {Zhang}, {Huang}, {Liu}, {Wang}, {Li}, and
  {Lin}]{Liu2021SPIE}
{Liu}, Y.; {Zhang}, X.; {Song}, T.; {Sun}, M.; {Liu}, D.; {Wang}, J.; {Zhao},
  M.; {Zhang}, T.; {Xu}, F.; {Fu}, H.;  et~al.
\newblock {Ground experiment of a 50~mm balloon-borne coronagraph for near
  space project}.
\newblock In \emph{Proceedings of the 10th International Symposium on Advanced
  Optical Manufacturing and Testing Technologies: Large Mirror and Telescopes};
  {Rao}, C.H., {Veillet}, C., {Ma}, X., {Fan}, B., {Liu}, F., {Collados Vera},
  M., Eds.;   {Society of Photo-Optical
  Instrumentation Engineers (SPIE) Conference Series}; SPIE: Bellingham, WA, USA, 2021; Volume~12070,  p. 120700B.
\newblock {\url{https://doi.org/10.1117/12.2605310}}.

\bibitem[{Li} et~al.(2024){Li}, {Huang}, {Zhou}, {Zhang}, and
  {Zhang}]{Li2024ChJSS}
{Li}, Y.; {Huang}, W.; {Zhou}, J.; {Zhang}, X.; {Zhang}, H.
\newblock {Development Status and Prospects of Near Space Observatories}.
\newblock {\em Chin. J. Space Sci.} {\bf 2024}, {\em
  44},~1068--1085.
\newblock {\url{https://doi.org/10.11728/cjss2024.06.2023-0145}}.

\bibitem[{Gopalswamy} et~al.(2021){Gopalswamy}, {Newmark}, {Yashiro},
  {M{\"a}kel{\"a}}, {Reginald}, {Thakur}, {Gong}, {Kim}, {Cho}, {Choi}, {Baek},
  {Bong}, {Yang}, {Park}, {Kim}, {Park}, {Lee}, {Kim}, and
  {Lim}]{Gopalswamy2021SoPh}
{Gopalswamy}, N.; {Newmark}, J.; {Yashiro}, S.; {M{\"a}kel{\"a}}, P.;
  {Reginald}, N.; {Thakur}, N.; {Gong}, Q.; {Kim}, Y.H.; {Cho}, K.S.; {Choi},
  S.H.;  et~al.
\newblock {The Balloon-Borne Investigation of Temperature and Speed of
  Electrons in the Corona (BITSE): Mission Description and Preliminary
  Results}.
\newblock {\em \solphys} {\bf 2021}, {\em 296},~15--46.
\newblock {\url{https://doi.org/10.1007/s11207-020-01751-8}}.

\bibitem[{Song} et~al.(2023){Song}, {Li}, {Zhou}, {Xia}, {Cheng}, and
  {Chen}]{Songa2023ApJ}
{Song}, H.; {Li}, L.; {Zhou}, Z.; {Xia}, L.; {Cheng}, X.; {Chen}, Y.
\newblock {The Structure of Coronal Mass Ejections Recorded by the
  K-Coronagraph at Mauna Loa Solar Observatory}.
\newblock {\em \apjl} {\bf 2023}, {\em 952},~L22--L28.
\newblock {\url{https://doi.org/10.3847/2041-8213/ace422}}.

\bibitem[{Liu} et~al.(2025){Liu}, {Yu}, {Zhang}, {Huang}, {Xia}, {Sun}, {Mao},
  {Sun}, {Tang}, {Fu}, {Liu}, {Zhang}, and {Han}]{Liu2025RAA}
{Liu}, D.Y.; {Yu}, X.Y.; {Zhang}, H.X.; {Huang}, Z.H.; {Xia}, L.D.; {Sun},
  M.Z.; {Mao}, X.L.; {Sun}, B.Y.; {Tang}, N.; {Fu}, H.;  et~al.
\newblock {Study on Real-time Monitoring Method for Dust-scattered Stray Light
  in the Spectral Imaging CoronaGraph of the Chinese Meridian Project Phase
  II}.
\newblock {\em Res. Astron. Astrophys.} {\bf 2025}, {\em
  25},~015014.
\newblock {\url{https://doi.org/10.1088/1674-4527/ad9a34}}.

\bibitem[{Tomczyk} et~al.(2008){Tomczyk}, {Card}, {Darnell}, {Elmore}, {Lull},
  {Nelson}, {Streander}, {Burkepile}, {Casini}, and {Judge}]{Tomczyk2008SoPh}
{Tomczyk}, S.; {Card}, G.L.; {Darnell}, T.; {Elmore}, D.F.; {Lull}, R.;
  {Nelson}, P.G.; {Streander}, K.V.; {Burkepile}, J.; {Casini}, R.; {Judge},
  P.G.
\newblock {An Instrument to Measure Coronal Emission Line Polarization}.
\newblock {\em \solphys} {\bf 2008}, {\em 247},~411--428.
\newblock {\url{https://doi.org/10.1007/s11207-007-9103-6}}.

\bibitem[{Morton} et~al.(2016){Morton}, {Tomczyk}, and {Pinto}]{Morton2016ApJ}
{Morton}, R.J.; {Tomczyk}, S.; {Pinto}, R.F.
\newblock {A Global View of Velocity Fluctuations in the Corona below 1.3 R
  $_{{\ensuremath{\odot}}}$ with CoMP}.
\newblock {\em \apj} {\bf 2016}, {\em 828},~89--101.
\newblock {\url{https://doi.org/10.3847/0004-637X/828/2/89}}.

\bibitem[{Yang} et~al.(2026){Yang}, {Rempel}, {Gibson}, and {de
  Toma}]{Yang2026ApJS}
{Yang}, Z.; {Rempel}, M.; {Gibson}, S.; {de Toma}, G.
\newblock {Measuring the Coronal Magnetic Field with 2D Coronal Seismology: A
  Forward-modeling Validation}.
\newblock {\em Astrophys. J. Suppl. Ser.} {\bf 2026}, {\em 283},~3--15.
\newblock {\url{https://doi.org/10.3847/1538-4365/ae39c5}}.

\bibitem[{Li} et~al.(2017){Li}, {Landi Degl'Innocenti}, and {Qu}]{Li2017ApJ}
{Li}, H.; {Landi Degl'Innocenti}, E.; {Qu}, Z.
\newblock {Polarization of Coronal Forbidden Lines}.
\newblock {\em \apj} {\bf 2017}, {\em 838},~69.
\newblock {\url{https://doi.org/10.3847/1538-4357/aa6625}}.

\bibitem[{Priyal} et~al.(2025){Priyal}, {Ramesh}, {Singh}, {Sasikumar Raja},
  and {Gopalswamy}]{Priyal2025ApJ}
{Priyal}, V.M.; {Ramesh}, R.; {Singh}, J.; {Sasikumar Raja}, K.; {Gopalswamy},
  N.
\newblock {Near-Sun Characteristics of a CME Inferred from Observations in the
  5303 {\r{A}} Coronal Emission Line}.
\newblock {\em \apj} {\bf 2025}, {\em 994},~182.
\newblock {\url{https://doi.org/10.3847/1538-4357/ae1a47}}.

\bibitem[{Tan} et~al.(2002){Tan}, {Cen}, {Qian}, and {Wang}]{Tan2002BASI}
{Tan}, H.; {Cen}, X.F.; {Qian}, T.L.; {Wang}, J.C.
\newblock {Evaluation of Lijiang Gaomeigu site for astrophysical observation}.
\newblock {\em Bull. Astron. Soc. India} {\bf 2002}, {\em
  30},~881--893.

\bibitem[{ASCOM Initiative}(2025)]{ASCOM}
{ASCOM Initiative}.
\newblock ASCOM Standards. 2025.
\newblock Available online:  \url{https://www.ascom-standards.org}, 
\newblock (accessed on 30 December~2025).

\bibitem[{Sha} et~al.(2023){Sha}, {Liu}, {Zhang}, and {Song}]{Sha2023SoPh}
{Sha}, F.; {Liu}, Y.; {Zhang}, X.; {Song}, T.
\newblock {Characterization and Correction of the Scattering Background
  Produced by Dust on the Objective Lens of the Lijiang 10-cm Coronagraph}.
\newblock {\em \solphys} {\bf 2023}, {\em 298},~139-154.
\newblock {\url{https://doi.org/10.1007/s11207-023-02233-3}}.

\bibitem[{Sha} et~al.(2025){Sha}, {Liu}, {Xia}, {Chen}, {Zhou}, {Chen},
  {Zhong}, {Zhang}, {Song}, {Sun}, {Yu}, {Li}, {Oloketuyi}, {Liu}, {Wang},
  {Luo}, and {Li}]{sha2025ApJ}
{Sha}, F.; {Liu}, Y.; {Xia}, L.; {Chen}, Y.; {Zhou}, Q.; {Chen}, Y.; {Zhong},
  C.; {Zhang}, X.; {Song}, T.; {Sun}, M.;  et~al.
\newblock {Mapping Ground-based Coronagraphic Images to
  Helioprojective-Cartesian Coordinate System by Image Registration}.
\newblock {\em \apj} {\bf 2025}, {\em 990},~56-65.
\newblock {\url{https://doi.org/10.3847/1538-4357/adf05e}}.

\bibitem[{Zhang} et~al.(2022){Zhang}, {Liu}, {Zhao}, {Song}, {Wang}, {Li}, and
  {Li}]{Zhang2022RAA007Z}
{Zhang}, X.F.; {Liu}, Y.; {Zhao}, M.Y.; {Song}, T.F.; {Wang}, J.X.; {Li}, X.B.;
  {Li}, Z.H.
\newblock {On the Relation Between Coronal Green Line Brightness and Magnetic
  Fields Intensity}.
\newblock {\em Res. Astron. Astrophys.} {\bf 2022}, {\em
  22},~075007.
\newblock {\url{https://doi.org/10.1088/1674-4527/ac6fb8}}.

\bibitem[{Squire} et~al.(2022){Squire}, {Meyrand}, {Kunz}, {Arzamasskiy},
  {Schekochihin}, and {Quataert}]{Squire2022NatAs}
{Squire}, J.; {Meyrand}, R.; {Kunz}, M.W.; {Arzamasskiy}, L.; {Schekochihin},
  A.A.; {Quataert}, E.
\newblock {High-frequency heating of the solar wind triggered by low-frequency
  turbulence}.
\newblock {\em Nat. Astron.} {\bf 2022}, {\em 6},~715--728.
\newblock {\url{https://doi.org/10.1038/s41550-022-01624-z}}.

\bibitem[{Linker} et~al.(2024){Linker}, {Downs}, {Caplan}, {Mason}, {Riley},
  {Palmerio}, {Ben-Nun}, {Davidson}, {Lionello}, {Reyes}, {Titov}, {Torok}, and
  {Turtle}]{Linker2024AGUF}
{Linker}, J.; {Downs}, C.; {Caplan}, R.M.; {Mason}, E.I.; {Riley}, P.;
  {Palmerio}, E.; {Ben-Nun}, M.; {Davidson}, R.; {Lionello}, R.; {Reyes}, A.;
  et~al.
\newblock {Prediction of the Structure of the Corona for the 2024 Total Solar
  Eclipse with a Near-Real Time Model}.
\newblock In \emph{Proceedings of the AGU Fall Meeting Abstracts}; American Geophysical Union: Washington, DC, USA, 9--13 December 2024;
  Volume 2024, p. SH51E--2943.

  \bibitem[{Linker} et~al.(2019){Linker}, {Downs}, {Caplan}, {Riley}, {Titov},
  {Lionello}, {Torok}, and {Reyes}]{Linker2019AGUF}
{Linker}, J.; {Downs}, C.; {Caplan}, R.M.; {Riley}, P.; {Titov}, V.S.;
  {Lionello}, R.; {Torok}, T.; {Reyes}, A.
\newblock {Prediction of Coronal Structure for the July 2, 2019 Total Solar
  Eclipse: Comparison with Observations}.
\newblock In \emph{Proceedings of the AGU Fall Meeting Abstracts}; American Geophysical Union: San Francisco, CA, USA, 9--13 December 2019; Volume 2019, p. SH13A--04.

\bibitem[{Zhang} et~al.(2022){Zhang}, {Liu}, {Zhao}, {Liu}, {Elmhamdi}, {Song},
  {Li}, {Li}, {Sha}, {Wang}, {Li}, {Shen}, {Liu}, {Liang}, and
  {Al-Shammari}]{Zhang2022RAA012Z}
{Zhang}, X.F.; {Liu}, Y.; {Zhao}, M.Y.; {Liu}, J.H.; {Elmhamdi}, A.; {Song},
  T.F.; {Li}, Z.H.; {Li}, H.B.; {Sha}, F.Y.; {Wang}, J.X.;  et~al.
\newblock {Comparison of the Coronal Green-line Intensities with the EUV
  Measurements from SDO/AIA}.
\newblock {\em Res. Astron. Astrophys.} {\bf 2022}, {\em
  22},~075012.
\newblock {\url{https://doi.org/10.1088/1674-4527/ac712e}}.

\bibitem[{Woods} et~al.(2024){Woods}, {Eden}, {Eparvier}, {Jones}, {Woodraska},
  {Chamberlin}, and {Machol}]{2024JGRA}
{Woods}, T.N.; {Eden}, T.; {Eparvier}, F.G.; {Jones}, A.R.; {Woodraska}, D.L.;
  {Chamberlin}, P.C.; {Machol}, J.L.
\newblock {GOES-R Series X-Ray Sensor (XRS): 1. Design and Pre-Flight
  Calibration}.
\newblock {\em J. Geophys. Res. (Space Phys.)} {\bf 2024},
  {\em 129},~2024JA032925.
\newblock {\url{https://doi.org/10.1029/2024JA032925}}.

\end{thebibliography}
\end{document}